\documentclass[fleqn,usenatbib]{mnras}

\usepackage{newtxtext,newtxmath}

\usepackage[T1]{fontenc}
\usepackage{ae,aecompl}

\usepackage{graphicx}	
\usepackage{amsmath}	
\usepackage{footnote}
\usepackage{longtable}

\usepackage[authoryear]{natbib}
\bibpunct{(}{)}{;}{a}{}{,}
\label{Bibliography}

\title[Searching for energy-resolved QPOs in AGN]{Searching for energy-resolved quasi-periodic oscillations in AGN}

\author[D. I. Ashton et al.]{
Dominic I. Ashton,$^{1}$\thanks{E-mail: d.i.ashton@soton.ac.uk}
\& Matthew J. Middleton $^{1}$ 
\\
$^{1}$Department of Physics \& Astronomy, University of Southampton, Southampton, SO17 1BJ, UK
\\
}

\date{Accepted XXX. Received YYY; in original form ZZZ}

\pubyear{2020}

\begin{document}
\label{firstpage}
\pagerange{\pageref{firstpage}--\pageref{lastpage}}
\maketitle

\begin{abstract}
X-ray quasi-periodic oscillations (QPOs) in AGN allow us to probe and understand the nature of accretion in highly curved space-time, yet the most robust form of detection (i.e. repeat detections over multiple observations) has been limited to a single source to-date, with only tentative claims of single observation detections in several others. The association of those established AGN QPOs with a specific spectral component has motivated us to search the \textit{XMM-Newton} archive and analyse the energy-resolved lightcurves of 38 bright AGN. We apply a conservative false alarm testing routine folding in the uncertainty and covariance of the underlying broad-band noise. We also explore the impact of red-noise leak and the assumption of various different forms (power-law, broken power-law and lorentzians) for the underlying broad-band noise. In this initial study, we report QPO candidates in 6 AGN (7 including one tentative detection in MRK~766) from our sample of 38, which tend to be found at characteristic energies and, in four cases, at the same frequency across at least two observations, indicating they are highly unlikely to be spurious in nature.
\end{abstract}




\section{Introduction}

The accretion of gas onto black holes provides insights into the behaviour of matter in the strong gravity regime and can therefore test general relativity (GR) itself. As the X-rays trace the inner-most regions (the disc and corona in black hole binaries -- BHBs, and the corona in AGN) it is the rapid variability in this band which encodes the most valuable diagnostic information. Such studies have proven capable of disentangling the geometry of the flow \citep{KLLinesin1H0707, Kara2019, Wilkins2013,2020Alston} whilst quasi-periodic oscillations (QPOs), ubiquitous in BHBs \citep{2016BHBQPOs} can provide information about the dynamics of the disc and the effects of external torques (e.g. the Lense-Thirring effect: \citealt{StellaVietri1999,Fragile2007,2011IngramDone}). 

Discovering QPOs in AGN has been of importance for some time, as they provide tests for the scalability of the accretion flow by black hole mass (e.g. \citealt{2006McHardyAGNScaling}), and a `slow-motion' view of the process creating them (i.e. many more photons per QPO phase bin relative to BHBs). QPOs detected in BHBs (with masses $\sim$ 10M$_{\sun}$) range from 1-100~s (the low frequency QPOs - LFQPOs), to 0.01 - 0.1~s (the high frequency QPOs - HFQPOs, see \citealt{Remillard2006XRBproperties}). Assuming a linear scaling to SMBH masses of $\sim$10$^{6-7}$M$_{\sun}$, implies AGN QPOs should have periods of days to weeks for LFQPO analogues, and hundreds of seconds to hours for the HFQPOs. It is therefore reasonable to speculate that, should a simple scaling hold, the latter should be present within some long observations of AGN taken by existing X-ray instruments.

To-date there have been numerous claims of X-ray QPOs in AGN, although only a very small number have proven to be statistically robust upon detailed investigation. The first widely accepted detection was in the Narrow Line Seyfert 1 (NLS1), RE~J1034+396 \citep{2008Nature} with a QPO period of approximately one hour, thereby resembling a mass-scaled BHB HFQPO  \citep{2010MiddletonBHBAnalogy}. This QPO has been speculated to originate in the hardest energy spectral component, similar to the 67~Hz QPO of GRS 1915+105 \citep{2009MiddletonSoftExcess, 2011Middleton8Years}, and has remained detectable at around the same frequency across several observations (\citealt{2014Alston5}, \citealt{2020JinREJ}). The chance probability of making a false signal appear across multiple observations at the same frequency is extremely low, and so the repeat detection of the QPO in RE~J1034+396 lends strength to its identification as a bona-fide signal. In addition to RE~J1034+396, there has been a QPO detection claim in a single observation MS~2254.9-3712 \citep{2015AlstonMS2254} with a period of $\sim$ 2 hours, and reports of possible QPOs following different methodolgies in 1H~0707-495, MRK~766, IRAS~13224-3809 and ARK~564 (\citealt{2007IMcHardyArk564Lor}; \citealt{2016Pan1HQPO}; \citealt{Zhang2017MRK766}, \citealt{Zhang20181H}; \citealt{2019AlstonIRASvariability}) (as well as in tidal disruption events, e.g. \citealt{2019PasahmTDEs}). The confirmation of tentative QPO claims and the discovery of new AGN QPOs is important; the present sample size of robust detections is extremely small which hinders a rigorous analysis of their origin, and critiques of their similarities or differences when compared to BHB QPOs.

Central to the detection of a QPO is a statistical method which robustly tests the null hypothesis, and accounts for the shape of the underlying broad-band noise onto which the QPO is imprinted \citep{2005Vaughan, 2010BayesianVaughan, 2016FalsePeriodicities}. Previous large-scale efforts include \citet{2012GonzMartinVaughan} who studied a large sample of 104 AGN and described the properties of the high frequency noise {\it en-masse} but without locating new QPOs. Here we build on previous work by searching for QPOs in {\it energy-resolved} AGN lightcurves, and present the discovery of several new statistically significant QPO candidates. 

The paper is laid out as follows. In Section 2 we introduce the AGN sample and our methods of data reduction for these observations. In Section 3 we outline the statistical framework for our QPO search, including the tests we apply to probe any candidates in greater detail. In Section 4 we outline the results of the QPO search. In Section 5 we discuss the additional uncertainty in the broad-band noise and its possible impact on QPO detections. Lastly, we discuss the relevance and impact of our findings in Section 6.

\section{Observations}

\subsection{AGN Sample}

We construct our AGN sample primarily from the Palomar Green QSO (PGQSO) catalogue used by \citet{2006CrummyCatalogue} and \citet{2007MiddletonPGQSRs} as these are extremely bright sources and have been well-studied by \textit{XMM-Newton}; this provides high signal-to-noise ratio data and the possibility of long (up to 130~ks) unbroken observations. In addition to this sample, we include 1H~0707-495, MS~22549-3712, NGC~4151, PHL~1092 and data from the most recent campaign of IRAS~13224-3809 (see \citealt{2019AlstonIRASvariability, 2020AlstonIRASreverbmap} for details). The total sample consists of 22 NLS1s and 16 Type-1 Seyferts, as defined by \citet{2010VernonCetty}. Of particular interest are the Narrow Line Seyfert 1 AGN, which include the widely-accepted case of X-ray QPOs in RE~J1034+396 \citep{2008Nature} as well as the reported case in MS 2254.9-3712 \citep{2015AlstonMS2254}.

\subsection{Data Reduction}

For each AGN in our sample, we considered all observations in the \textit{XMM-Newton} public HEASARC archive\footnotemark\footnotetext{https://heasarc.gsfc.nasa.gov/} up to September 2019 (see online table). We selected EPIC-PN data only for this analysis, as the PN has a higher throughout than the MOS detectors leading to a lower Poisson noise contribution. A circular aperture source region of 40" radius was centred on each AGN, with a background source region (also a circular aperture of 40" radius) positioned on the same CCD, but away from the source and chip read-out direction. We then followed standard procedure of selecting only single/double pixel events (with \texttt{PATTERN<=4}). Using the standard \textit{XMM-Newton} data reduction pipeline \footnote{https://www.cosmos.esa.int/web/xmm-newton/sas-thread-timing}, we extracted source and background light curves, using \texttt{EVSELECT} in \textit{SAS \textnormal{v}17.0.0}. We extracted lightcurves with a binsize of 100~s and utilised a sliding energy window across the nominal 0.3 - 10~keV range of the instrument. This window is split into 50 approximately linear energy bins, allowing 1225 energy combinations to be studied (e.g. 0.3 - 0.5 keV, 0.3 - 0.7 keV etc.). We proceeded to use \texttt{EPICLCCORR} to subtract the background and apply the necessary corrections to the light curves. Whilst detector pile-up is present in some of the observations, the redistribution of counts from soft to hard energies (where the count rate is typically much lower) is not expected to lead to false QPO detections as non-QPO counts act to dilute the signal's presence. Pile-up could, however, result in QPO candidates being located in energy bands in which they are not intrinsically the strongest. We will re-visit this issue in a forthcoming paper and here re-iterate that there will be some unaccounted-for distortion.

For each observation, the high-energy (10 - 12~keV), full-field background lightcurve was extracted with a binsize of 10s in order to identify soft-proton events which could contaminate the source lightcurve. We apply a cut-off such that any flares at 5$\sigma$ (where $\sigma$ is the standard deviation in the 10-12~keV lightcurve) or more above the stable (non-flaring) count rate level are removed. This process can introduce gaps into the lightcurves, which, if left unmodified, have the potential to introduce peaks in power (making QPO detections harder). Whist there is no standard method to correct for gaps, previous approaches include linearly interpolating between short missing segments of data (e.g. $<$ 200s), with count-rate variations accounted for by drawing from a Poisson distribution \citep{2012GonzMartinVaughan, 2015AlstonMS2254}. This method has the caveat of introducing a non-red noise background component to the light curve, which cannot replicate the component which has been removed, and thus is best applied sparingly (e.g. the interpolation used by \citealt{2015AlstonMS2254} is typically applied to < 0.5\% of the duration of the lightcurve). 

\begin{figure*}
    \centering
    \includegraphics[width=\textwidth]{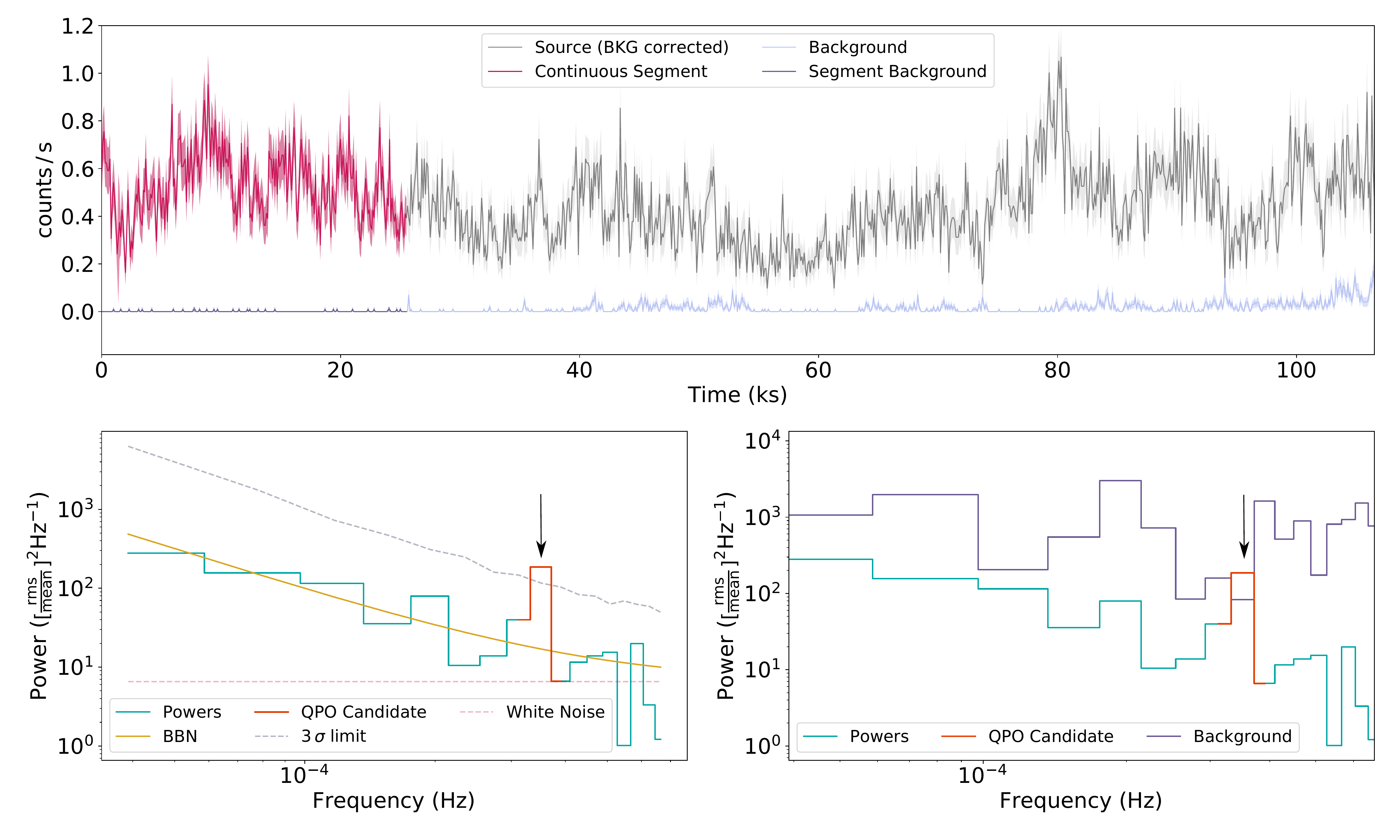}
    \caption{Top: the entire {\it XMM-Newton} EPIC-PN light curve for NGC 4051, OBS ID: 0109141401, with the longest continuous (flare-less) segment highlighted. Bottom left: periodogram of the background subtracted, flareless segment, with best fitting {\sc plc} model, simulated $3\sigma$ line and QPO candidate highlighted. Bottom right: a direct comparison between the source and background periodogram, highlighting the lack of a corresponding peak of power in the background at the QPO candidate frequency.}
    \label{fig:Triple_plot_NGC}
\end{figure*}

In our analysis, we take the most conservative approach of considering only the longest continuous segment between flares (see Figure \ref{fig:Triple_plot_NGC}). This avoids all interpolation assumptions, but has the limitation of reducing the amount of data available for analysis. However, we find that many of the longest continuous segments in our sample are of sufficient duration to probe down to $\le 10^{-4}$~Hz, below the frequencies of previously detected AGN QPOs \citep{2008Nature, 2015AlstonMS2254}.

We also note that we automatically create the lightcurve of the background, the periodogram of which we can inspect to ensure that any potential QPO candidate is associated with the AGN. An example of this is shown in Figure \ref{fig:Triple_plot_NGC}, where the absence of a feature in the background confirms the QPO candidate originates in the AGN, in this case, in NGC~4051.

\section{Periodogram analysis and QPO search}

A robust statistical framework in which to search for QPOs in broad-band noise is vital (see \citealt{Vaughan05WhereAretheQPOs?}), and we base our approach on that of \citet{2005Vaughan}. We produce periodograms for each energy-resolved lightcurve (reduced to $2^n$ bins where $n$ is an integer) by applying a fast-Fourier transform (FFT) and a fractional-rms normalisation \citep{2003UttleyVaughan}. Using the periodogram (rather than averaged power spectral distribution: PSD) allows us to maximise the number of Fourier frequencies, improving our constraints when modelling, whilst also allowing for QPO searches down to the lowest possible frequencies in the lightcurve. 

Similar to the work by \citet{2012GonzMartinVaughan}, we fit the periodogram using the maximum likelihood statistic. Following \citet{2005Vaughan}, we see that maximising the likelihood, $\mathcal{L}$, is equivalent to minimising $S \equiv -2 \ln{[\mathcal{L}]}$. Assuming a model $P_j(\hat{\theta}_{\rm k})$, for parameters $\hat{\theta}_{\rm k} = (\hat{\theta}_1 ... \hspace{0.065cm} \hat{\theta}_{\rm M})$, observed powers $I_{\rm j}$, and frequencies j$ = 1,2,...,n-1$, $S$ may be written as:

\begin{equation}
S = 2 \sum_{j=1}^{n-1} \left\{ \ln[{P}_{\rm j}] +\frac{I_{\rm j}}{{P}_{\rm j}} \right\}
\end{equation}

We chose to apply two models to our data, the first ({\sc plc}) is a power law plus constant:

\begin{equation}
P(\nu) = N_0 \nu^{\hspace{0.065cm}\beta} + C
\end{equation}

\noindent and consists of three free parameters: $\beta$, the spectral index, $N_0$, the normalisation and $C$, a constant to account for Poisson noise (white noise). The second model ({\sc bknplc}) takes the form of a broken power-law, with break frequency, $\nu_b$:

\[P_1(\nu) = N_1 \nu^{\hspace{0.065cm}\beta_1} + C \hspace{0.5cm} \textnormal{for } \nu \leq \nu_{\rm b} \]
\vspace{-0.5cm}
\begin{equation}
P_2(\nu) = N_2 \nu^{\hspace{0.065cm}\beta_2} + C \hspace{0.5cm} \textnormal{for } \nu > \nu_{\rm b}
\end{equation}

\noindent At frequencies at, or below the break ($\nu \leq \nu_{\rm b}$) we assume a spectral index $\beta_1$ and normalisation $N_1$, and at frequencies above the break ($\nu > \nu_{\rm b}$), we assume a spectral index $\beta_2$. The normalisation above the break, $N_2$, is calculated directly from the previous parameters. In both models we set the constant ($C$) to the predicted level of the white noise following the method described in \citet{2003UttleyVaughan}:

\begin{equation}
C = \frac{2 dt}{\langle x \rangle ^2} \Bar{\sigma}^2 
\end{equation}

\noindent with timebin size, $dt$, mean count rate, $ \langle x \rangle$ and mean square count rate error $\Bar{\sigma}^2$ of the background subtracted lightcurve.

\subsection{Analysis Procedure}

We first apply the {\sc plc} model to our periodogram, using maximum likelihood statistics inbuilt into a limited memory Broyden-Fletcher-Goldfarb-Shanno (L-BFGS-B) minimisation algorithm, with bounds set at (-10, 10) for $\beta$ and (0, $\infty$) for the normalisation, $N_0$. We perform a preliminary fit with this minimisation routine to obtain the parameters $(\hat{\beta}, \hat{N_0})$ which are successful in minimising the negative log likelihood, $S$. Following the procedure of \citet{2005Vaughan}, we perform a frequency cut where the Poisson noise begins to dominate over the red noise (where $C > P_{\rm j}$), and ignore frequencies above this. We set a minimum number of Fourier frequencies in the remaining periodogram to ensure we have a sufficient number of degrees of freedom to constrain the underlying noise with simple models (in practice we have set an arbitrary lower limit of 13 Fourier bins). Note, we reject any lightcurve/energy-bin where the fitting returns a positive $\beta$, although this only occurs in the case of poor quality or Poisson noise-dominated data.

\subsubsection{Defining the null hypothesis}

We define a QPO candidate to be a peak of power above the broad-band noise, and for simplicity, that this lies within a single Fourier frequency bin. We follow \citet{2005Vaughan}, where we compute and then maximise the ratio $R = \frac{2I_{\rm j}}{P_{\rm j}}$, comparing the observed powers $I_{\rm j}$ and the model powers, $P_{\rm j}$, across all frequencies. The frequency, $j$, in which $R_{\rm j}$ is maximised, is defined as the QPO candidate frequency, $\nu_{\rm Q}$, with power, $P_{\rm Q}$. 

In order to determine the statistical significance of any outlier (or QPO candidate), we must first define our null hypothesis. We define our null hypothesis to be the power in the QPO candidate, $P_{\rm Q}$, being a random fluctuation of the underlying red noise, intrinsic to the AGN. Following \citet{2005Vaughan}, we remove the candidate from the periodogram (leaving $\ge$ 12 bins as per our restrictions mentioned above) and refit up to the previously defined white noise cutoff frequency, with a {\sc plc} model in order to obtain the best description of the underlying continuum (the impact of fixing the position of the white noise cutoff is discussed in section 5.1). We obtain constraints on the best-fitting model parameters by applying a bootstrap-with-replacement L-BFGS-B minimisation algorithm, using maximum likelihood statistics over 10,000 iterations. This bootstrap-with-replacement method has several advantages over a single fitting iteration including reducing the significance of any individual bin variance.

For each iteration of the bootstrap, we randomly draw initial parameter estimates for use within the minimisation algorithm, using an arbitrary Gaussian distribution centred on the parameters of the initial periodogram fit -- $(\hat{\beta}, \hat{N_0})$ -- and with a standard deviation of $\sim 10\%$. This simple measure ensures the minimisation algorithm most effectively searches across parameter space for a true global minimum and does not prematurely meet the minimisation criteria.

Over the 10,000 bootstrap iterations, we require a minimum of 5$\%$ to fulfil the minimisation success criteria, which includes returning a negative value of $\beta$. We group the resulting distribution of successfully minimised bootstrapped parameters into 100 bins and take the 1$\sigma$ distribution about the 50th percentile, to constrain variation in mode. The modal values of these distributions represent our new minimised set of parameters $(\hat{\beta}, \hat{N_0})$ which describe the {\sc plc} of the broad-band noise minus the QPO candidate. 

We note that in restricting our search to excess power residing at a single Fourier frequency, any lower coherence feature which spans more than one bin will lead to us inferring a flatter power-law index and higher normalisation for the underlying broad band noise. We act to reduce the impact of this through our bootstrap fitting approach, which minimises the effect of individual bins on the overall model fit.

\subsubsection{Power-law vs broken power-law}

Given sufficient data quality, AGN power spectra spanning a large frequency range may be better described by a broken power law \citep{2002UttleyBBN, 2007IMcHardyArk564Lor,2012GonzMartinVaughan}. To explore the presence of a break, we also fitted each periodogram with a {\sc bknplc}, and used a Bayesian Information Criterion (BIC) test \citep{Liddle2007} to evaluate the statistical preference between the {\sc plc} and {\sc bknplc} models. For the fitting process, we again used an L-BFGS-B minimisation algorithm, with bounds of (-10, 0), (-10, 0), (0, $\infty$) for parameters $\beta_1, \beta_2$ and $N_1$ respectively, over a set of possible break frequencies: $\nu_{\rm b,j}$ (we note that we require $\beta_1$ and $\beta_2$ to only take physically relevant, negative values). In order to perform a consistent BIC test, we kept the number of frequency bins the same between {\sc plc} and {\sc bknplc} models, with possible break frequencies set to lie below the white noise cutoff frequency. Finally, we reject any maximum likelihood model fit where $\beta_2 > \beta_1$. The BIC is defined as: 
\begin{equation}
BIC = S + k \ln{(m)} 
\end{equation}
\noindent with negative log-likelihood, $S$, number of model parameters, $k$ and number of data points, $m$ (the number of data points in our periodogram). The $BIC$ value is minimised for the most statistically preferred model, i.e. it allows us to identify which model, {\sc plc} or {\sc bknplc}, best describes our observed data (we explore the impact of noise entering from outside of our bandpass in Section 5). 

Across our sample of non-white noise dominated observations (i.e. those with sufficient data points for our analysis), we find the {\sc bknplc} model is preferred in $\sim10\%$ of cases. By comparison \citet{2012GonzMartinVaughan}, found $16\%$ of AGN PSDs in their sample to be best described by a bending power-law, and, as with our analysis, the majority by a simple power-law. For simplicity, we chose to focus on the remaining datasets best described with a {\sc plc} model, although we note the possibility that -- should a break be present -- in some cases we will lack a sufficient number of frequency bins below it to constrain its presence. We explore the impact of such unreported breaks on our QPO candidate detection -- and the potential impact of more complex broad band noise models -- later in the paper.

\subsubsection{Uncertainty on the best-fitting {\sc plc} model}

In order to have a more complete grasp of our null hypothesis, we consider the uncertainty on the maximum likelihood parameters which describe the broad-band noise model. To do so we can follow the approach of \citet{2005Vaughan} and use $\Delta S = -2\Delta\ln{[\mathcal{L}]}$ to map confidence contours, analogous to $\Delta \chi^2$. For a generic M-dimensional model, we would expect the relationship between parameters to form an M-dimensional paraboloid. \citet{2005Vaughan} demonstrates this for the simple 2-dimensional power-law case, where the relationship between parameters $(\beta, \log[N_0])$ forms a highly covariant ellipse. The 1$\sigma$ parameter limits are defined following $\chi^{2}_{\nu}$ where in this case, $\nu$ is the number of model parameters (such that 1$\sigma$ for a 2-dimensional model occurs at $\Delta$S = 2.3), and may be determined from mapping the $\Delta S$ space. Given these confidence limits, it is then possible to determine the terms of the covariance matrix \citep{coe2009fisher} describing the M-dimensional Gaussian distribution, which can then be used as the basis for Monte-Carlo simulations.

A direct approach to modelling the parameter space is to use a numerical grid; this allows the 1$\sigma$ parameter limits to be accurately determined, but is computationally intensive and also dependent on the data quality (which sets the scale of the initial grid). Semi-analytical methods assuming a Gaussian symmetry in the parameter space can also provide reasonable limiting values, however, such symmetry does not hold universally. Our chosen bootstrap method (section 3.1.1) {\it automatically} constructs a discretised distribution of model parameters, centred on the best fitting values. This discrete distribution represents a true sampling of probability space, a distribution which we then seek to describe with a continuous function. To model the discretised bootstrap dataset and directly access the covariance matrix, we applied a Gaussian-mixture-model (GMM), unsupervised machine learning approach, which identifies and describes the data as a continuous model. GMM uses a single Gaussian, or sum of Gaussian distributions, with maximum likelihood statistics in its fitting process, with the covariance matrix defined internally. We use the GMM algorithm implemented within {\sc scikit-learn} \citep{scikit-learn}. 

\begin{figure*}
    \centering
    \includegraphics[width=0.8\textwidth]{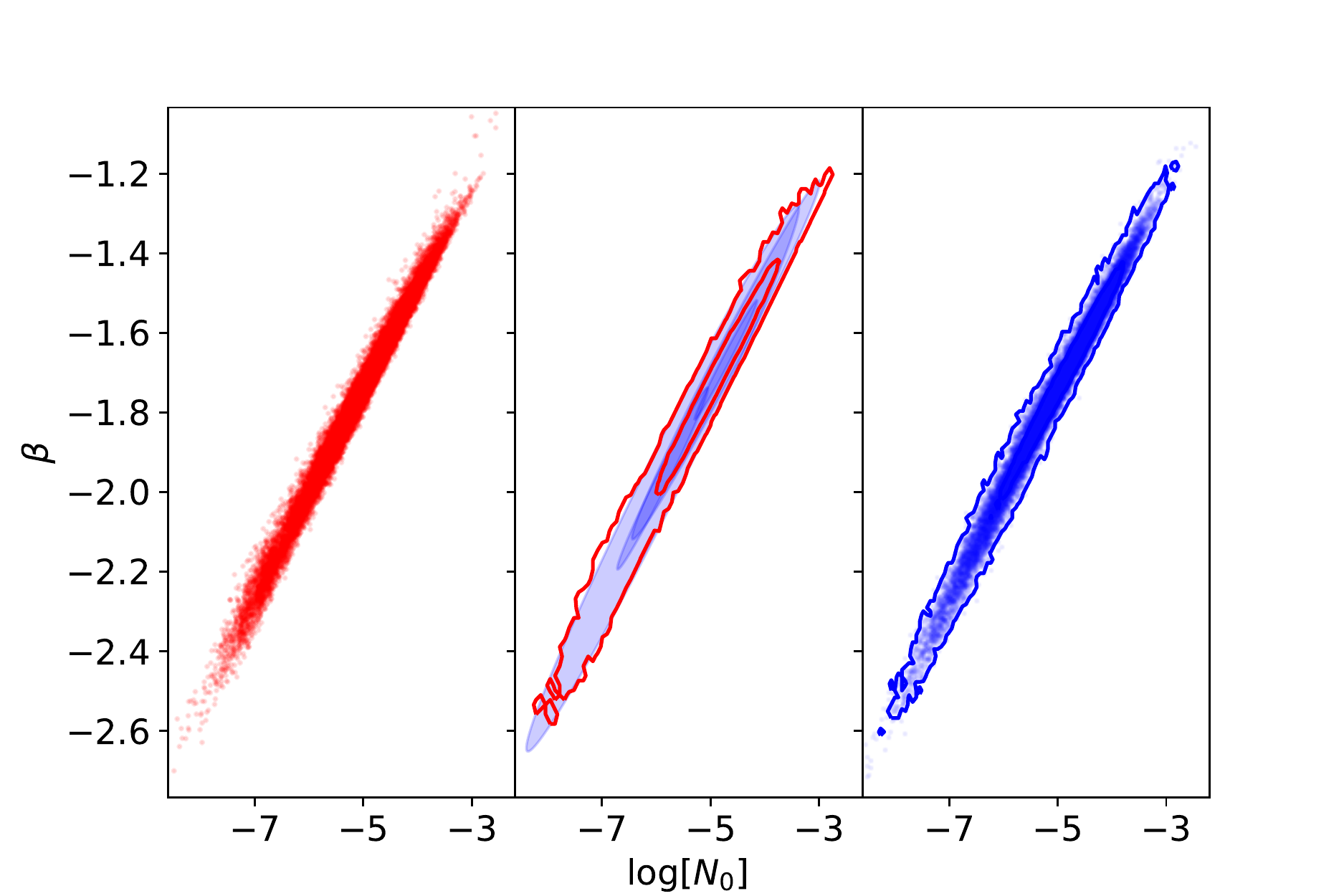}
    \caption{Plots showing our approach to modelling the parameter space of the broad-band noise, used to define our null hypothesis and constrain the QPO candidate significance. Left: discretised bootstrap distribution of {\sc plc} parameter space. Middle: a two-stage (i.e. two overlapped Gaussians) GMM model (blue), over bootstrap $\sim68\%$ and $\sim99.7\% $ probability density contours. Right: individual GMM parameter draws, with highlighted $\sim68\%$ and $\sim99.7\% $ probability density contours (i.e such that $\sim99.7\% $ of parameters drawn directly from the GMM lie within the outer contour).}
    \label{fig:GMM_plot}
\end{figure*}

In order to determine the number of independent Gaussians needed to best describe a given boostrapped parameter space, we again apply the BIC test. For every iteration of GMM, we allow the fitting of an additional Gaussian distribution, up to a maximum of 10 iterations. The BIC penalises the added complexity of additional model parameters and the $BIC$ value is minimised in the case where the likelihood is maximised. For each iteration, the $BIC$ value is recorded, with the ideal number of GMM parameters determined from the iteration in which the $BIC$ is minimised. This method most efficiently allows us to model the $\Delta S$ parameter space with a continuous model, an example of which is shown in Figure \ref{fig:GMM_plot}.

\subsubsection{Effect of low count rates}

\begin{figure}
    \centering
    \includegraphics[width=0.47\textwidth]{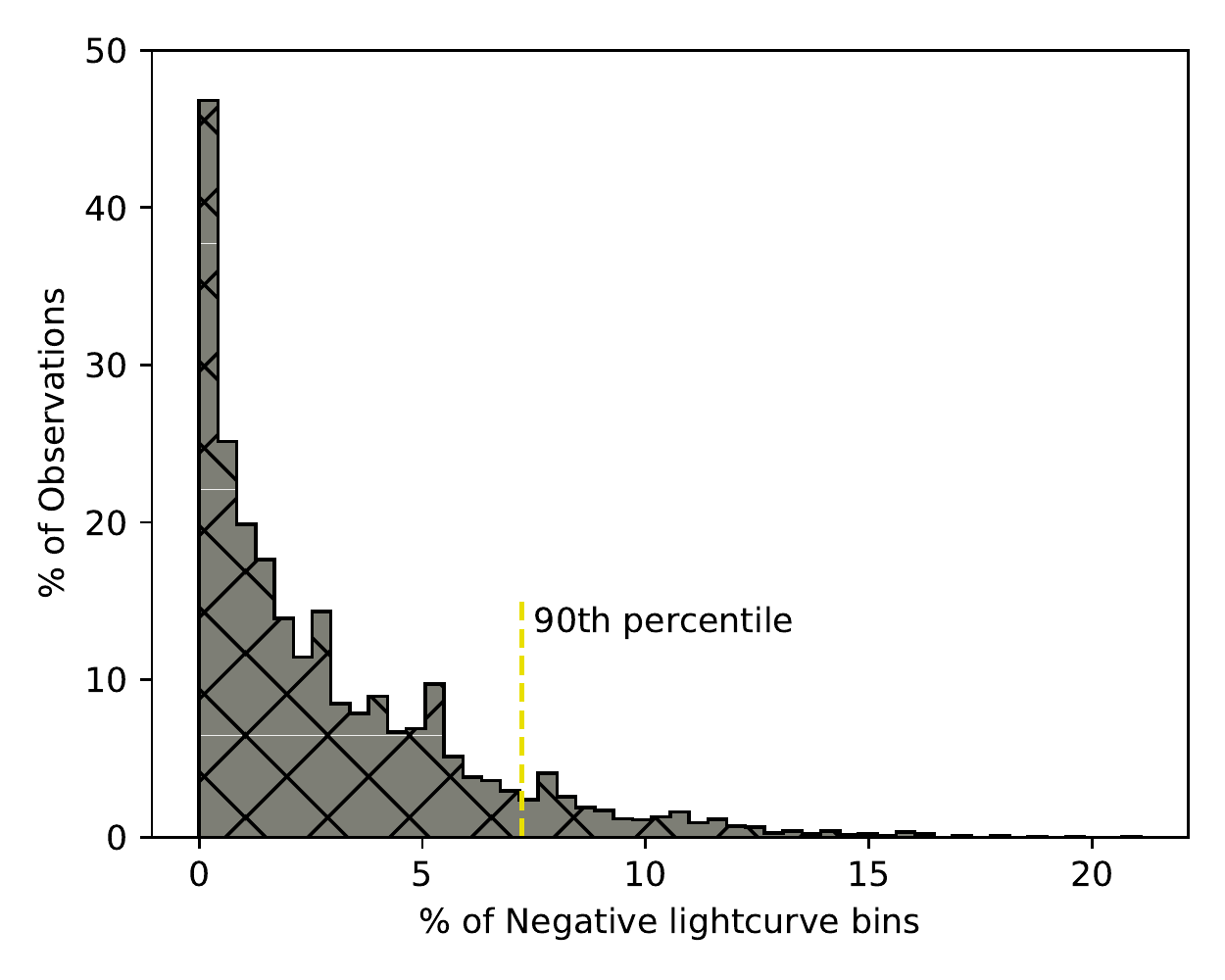}
    \caption{Histogram showing the percentage of negative count rate bins in our energy-resolved lightcurves with a minimum of a single negative bin ($\sim30\%$ of sample) and with periodograms determined to be above the white noise dominated threshold ($\ge 13$ frequency bins at frequencies below the cut-off). We find that most lightcurves have $<1\%$ negative bins, with the 90th percentile at $\sim7\%$.}
    \label{fig:Histogram_ncr}
\end{figure}

The combination of utilising narrow energy bands with falling count rates above the nominal response of the detector, can result in source count rates approaching the background count rate at high energies, even for the most luminous AGN (e.g. 1H~0707-495). Due to the stochastic (Poisson) nature of the background, in the case of very low source count rates, the background-subtracted lightcurve may contain bins with unphysical, negative count rates. Across our full sample of $\sim 240,000$ energy resolved lightcurves, we find a minimum of a single negative count rate bin in $\sim30 \%$ of cases. However, as shown in Figure \ref{fig:Histogram_ncr}, the vast majority of such lightcurves contain only a very small fraction of negative bins overall. For those cases where negative bins are present, we find the mean proportion of negative bins to be $\sim 3 \%$ of the lightcurve, with the $90$th percentile at $7\%$.

We explore the impact of negative count rates on our analysis, by simulating red noise lightcurves (using a {\sc plc} model and the method outlined in \citet{1995TK} and changing the fraction of negative count rates (by changing the mean count rate of the background). The bias introduced from such negative count rate bins can then be directly inferred from the difference between the input index and that determined from fitting the resulting periodogram ($\Delta \beta$ = $\beta_{\rm in} - \beta_{\rm out}$).

\begin{figure}
    \centering
    \includegraphics[width=0.47\textwidth]{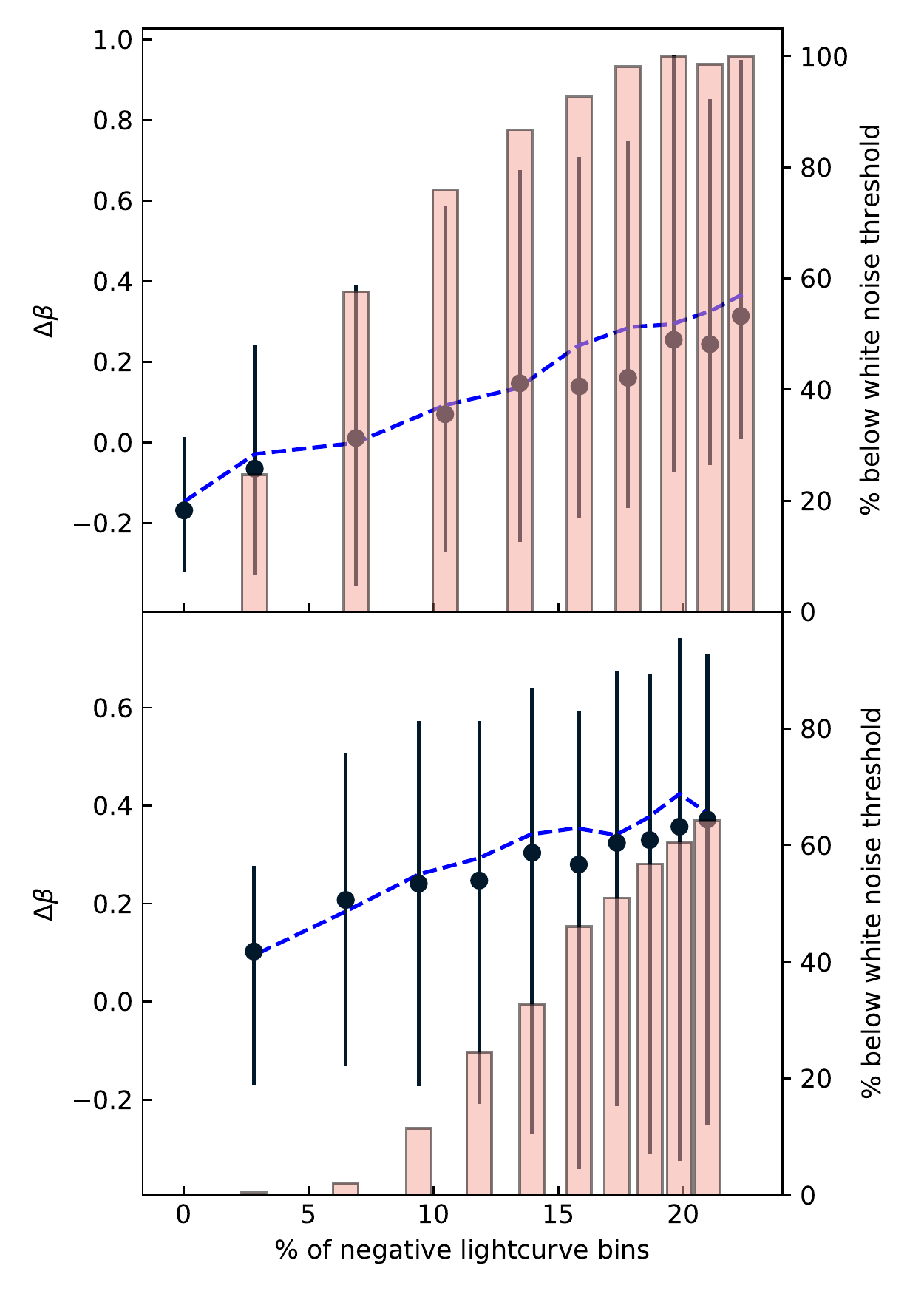}
    \caption{The impact on the inferred power spectral index, $\Delta\beta$, as a function of the percentage of negative bins found in a simulated Poisson-noise subtracted, red noise lightcurve (with 68\% confidence intervals plotted). The blue dashed line shows the observed trend when we instead replace these negatives with zeros. The bars correspond to the proportion of lightcurves where the periodogram would be rejected by our requirement to have 13 or more Fourier bins at frequencies below where the white noise starts to dominate. Top panel: example case where \textsc{PLC} index $\beta \sim -1$, Bottom panel: example case where \textsc{PLC} index $\beta \sim -2$.}
    \label{fig:Simulated_ncr}
\end{figure}

To perform the simulations, we utilised the best fitting models for IRAS~13224-3809, OBSID:0792180501 ($\beta = -1.95, \ln{N_0} = -5.09$) and ARK~564 OBSID:0670130801 ($\beta = -0.99, \ln{N_0} = -2.59$) as these both have \textsc{PLC} models which well describe the data, with indices close to $-2$ and $-1$ respectively, allowing us to explore \textsc{PLC} parameter dependence. From the lightcurves generated following \citep{1995TK}, we added Poisson noise to the lightcurves based on the corresponding mean count rate and a Poisson noise background to simulate X-ray background in the detector. To simulate a real observation, we then subtracted a new Poisson noise background, to leave the new background-subtracted lightcurve, the periodogram of which we model following the method already described. We repeat the simulation $10,000$ times in total (across a range of increasing background count rates) to sample the red noise, with the results shown in Figure \ref{fig:Simulated_ncr}. This indicates that for the vast majority of our actual lightcurves containing negative bins ($7\% \textnormal{ at the } 90$th quantile level), their presence introduces a bias error on the index of $\Delta \beta < 0.2$, which lies comfortably within the expected $68\%$ confidence intervals of $\beta$ values using a {\sc plc}, as seen in Figure \ref{fig:Simulated_ncr}. We find this effect is slightly more pronounced in cases of steeper indices, $\beta \sim -2$, but even at the limit of the \% of negative bins observed in our real data, this effect is negligible compared to the uncertainty from fitting with a \textsc{PLC} model itself -- as can be seen in Figure \ref{fig:GMM_plot}. We note that, for both \textsc{PLC} models, the number of simulated cases which our methodology would deem to be white noise dominated (where the number of Fourier bins below the white noise cut off is $<$ 13) increases with the \% of negative bins. This is especially true in simulated cases with flatter indices ($\beta \sim -1$), where this proportion becomes very high, at $>70\%$ of cases when considering $10\%$ negative bins. The application of our strict cut-off has helped avoid these issues and leaves us with typically only a very small proportion of negative bins which introduce only a small source of secondary error unlikely to impact our results.

We also explore an alternative method of replacing all negative bins with zeros (as shown in Figure \ref{fig:Simulated_ncr}), finding that this introduces an increased level of uncertainty in the index, with $\Delta \beta$ increasing linearly with the fraction of negative bins in the time series, similarly to the case where we keep negative bins. This approach clearly does not act to reduce the impact on the measured index, $\Delta \beta$, as we introduce a non-stochastic count rate which does not reflect the nature of the stochastic source and background. We therefore gain no benefit from this method and so avoid such an approach.

\subsubsection{QPO candidate significance testing}

To evaluate the significance of outliers identified in our energy-resolved periodograms, we simulated red noise power spectra, produced through the procedure detailed in \citet{1995TK} accounting for the small white noise contribution through the addition of the constant offset, $C$, which we use as an approximation for Poisson noise in our stochastic background explicitly -- as its contribution is small. For each candidate, we simulated 10,000 fake periodograms using a {\sc plc} model to describe the broad band noise, with parameters drawn at random from the mapped M-dimensional parameter space described above. This approach accounts for the intrinsic uncertainty in our estimated broad-band noise and therefore provides a more conservative significance estimate for a QPO candidate. We treated each simulated dataset in an identical manner to the real data, first by fitting each fake periodogram using a maximum likelihood minimisation routine (using a {\sc plc} model only), before identifying the largest oulier in each, removing it and refitting the underlying continuum as before. If the fitting routine was unsuccessful (as there is a low but finite chance that the fitting algorithm cannot minimise), we drew a new fake dataset, as we require the number of simulations to equal 10,000. In each successful case, we then evaluated the ratio $R_{\rm TK} = \frac{2I_{\rm TK}}{P_{\rm TK}}$ (where, as before, $I$ indicates the observed power, and $P$ the best-fitting model power at the same frequency). A false detection occurs if we find $R_{\rm TK} > R_{\rm Q}$, at any frequency (thereby treating each frequency as an independent free trial). Across $N$ simulations, we obtain $M$ datasets with false detections and the global significance, $GS$, of the outlier via:

\begin{equation}
GS = 1 - M/N
\end{equation}

\noindent To exceed a 3$\sigma$ significance, we require $GS > 0.9973$ (equivalent to $M <$ 27, given $N = 10,000$). Any outlier which meets this requirement is then considered to be a QPO candidiate - it is important to note that any such signal identified in isolation must still be treated with caution until additional evidence is obtained (e.g. the persistence of the signal across multiple observations and at similar frequencies -- \citealt{2014Alston5} -- or other characteristic behaviour which distinguishes it from the broad-band noise). In a similar fashion, we can produce $3\sigma$ confidence limits on observed powers from our Monte-Carlo simulations following the method of \citet{2005Vaughan} where the false alarm probability is determined at the candidate frequency and then corrected {\it post priori} for the number of free trials (typically the number of frequency bins in the periodogram). Although we use the global significance described above in identifying our QPO candidates, we also display these latter $3\sigma$ contours in our figures (e.g. Figure \ref{fig:Triple_plot_NGC}) for illustrative purposes. We also must be cautious of any detections close to the white noise cutoff, for at these frequencies, the relative fraction of Poisson noise (which we assume is constant rather than providing any additional stochastic variability) increases relative to the simulated red noise. 

Whilst we treat each frequency as independent, we do not consider each energy bin to be a free trial, as we expect some level of correlation between bands (as evidenced in AGN covariance spectra - e.g. \citealt{2011Middleton8Years}); we will explore techniques to address this directly in future. Finally, we note that one could take each AGN observation itself to be an independent free trial, however, as we do not yet know the conditions required to produce a QPO, nor for how long they persist, we take a more heuristic approach and treat only the frequencies as independent trials. 

\subsection{Further QPO candidate significance tests}

As we have suggested earlier, it is plausible that, regardless of the fact a {\sc plc} model may be statistically preferred in a given case, our BIC test may exclude a {\sc bknplc} model due to a low number of frequency bins below any putative break. Any unaccounted-for break in the broad-band noise has the potential to produce spurious peaks of power and so we perform a series of further tests to better judge the validity of any QPO candidates located with a {\sc plc} model. 

\begin{enumerate}
    \item In only those energy bands where a QPO candidate is located to $\ge$ 3$\sigma$, we refit the entire frequency bandpass (minus the {\it a priori} known QPO candidate frequency bin, but including the previously discarded white noise) with a {\sc bknplc} model. We allow the index above and below the break ($\beta_{1}, \beta_{2}$) and the break frequency $\nu_{\rm b}$ to be free (although in the case of the break set to be $\le \nu_{\rm Q}$). We establish the new white noise cutoff frequency -- which can differ from that found in the {\sc plc} fitting (see Section 5.1), and freeze the best fitting model (break frequency, $\beta_{1}$, $\beta_{2}$ and normalisation). We then simulate assuming this best-fitting model for the broad band noise and obtain a new GS value for the QPO candidate. 
    
    \item In only those energy bands where a QPO candidate is located to $\ge$ 3$\sigma$, we refit the entire frequency bandpass once again with a {\sc bknplc} model but with a fixed lower index, $\beta_{1}$ set to -1.1 \citep{2019AlstonIRASvariability} and with a break frequency $\nu_{\rm b} \le \nu_{\rm Q}$. As before, we freeze the best fitting model and obtain a new GS value for the QPO candidate.

We note that our requirement for $\nu_{\rm b} \le \nu_{\rm Q}$ is motivated by observations of black hole binaries (notably GRS~1915+105), where the high frequency QPOs (the natural  analogues to AGN QPOs at these frequencies \citep{2010MiddletonBHBAnalogy} occur above the high frequency break in the PSD \citep{1997MorganGRS1915}.

\end{enumerate}

\begin{table*}

\caption{$3 \sigma$ QPO candidate summary. The columns indicate: (1) Observation ID, (2): The longest continuous segment duration of the light curve, (3): the highest value of global significance obtained in a single energy bin, (4): the energy range of this bin, (5): the Fourier frequency of the QPO candidate detected at $\ge 3\sigma$, (6): the Fourier frequency bin width, (7): the (peak significance) fractional rms (FRMS) value at the QPO candidate Fourier frequency, (8): the QPO candidate Quality factor ($Q = \frac{\nu}{\Delta\nu}$ which we point out is not the product of the duration and QPO candidate frequency due to restricting our lightcurve to 2$^{n}$ bins). \textbf{*} We note that IRAS~ 13224-3809, OBSID:0792180201, 1H~0707-495 OBSID: 0653510601 and MRK~766, OBSID:0304030301 are considered borderline $3 \sigma$ results with ambiguous detections assuming a \textsc{plc} model. \textbf{$\dagger$} highlighted observations are those which are not validated by our further tests (see Section 3.2). All SMBH masses are provided in the form: $\log_{10}(\frac{M}{M_{\odot}})$ 
The estimated SMBH masses are obtained from: 
 \textsuperscript{\textit{b}} \citet{Done2012}, 
 \textsuperscript{\textit{c}} \citet{2019AlstonIRASvariability}, 
 \textsuperscript{\textit{d}} \citet{2016DoneJin}, 
 \textsuperscript{\textit{e}} \citet{2004GierlinskiDone}, \textsuperscript{\textit{f}} \citet{Denney2010}, 
 \textsuperscript{\textit{g}} \citet{Wang2001}, 
 \textsuperscript{\textit{h}} \citet{Giacch2014}.
}

\bigskip
\begin{tabular*}{\textwidth}{c @{\extracolsep{\fill}} ccccccc}
  \hline
  Obs ID & Duration (ks) & Peak &  Peak & QPO candidate & Frequency & FRMS & Q\\
  &&Significance (GS) & Energy (keV) & Frequency (Hz) & Error (Hz) & Value \\
 
 \hline
 \textbf{RE~J1034+396, \ $M_{\rm BH} \sim 6.17$}\, \textsuperscript{\textit{a}} \textsuperscript{\textit{b}} \\ \hline

 0506440101 & 28.94 & 1.0000 & 0.7 - 1.0 & $2.34\times10^{-4}$ & $3.91\times10^{-5}$ & 0.071 & 5.98 \\
 0655310201 & 41.60 & 1.0000 & 1.2 - 2.4 & $2.34\times10^{-4}$ & $3.91\times10^{-5}$ & 0.123 & 5.98 \\
 0675440101 & 23.92 & 1.0000 & 0.7 - 3.4 & $2.34\times10^{-4}$ & $7.81\times10^{-5}$ & 0.097 & 3.00 \\
 0675440201 & 28.82 & 1.0000 & 0.9 - 6.2 & $2.73\times10^{-4}$ & $3.91\times10^{-5}$ & 0.112 & 5.98 \\

 \hline
 \textbf{IRAS 13224-3809, \ $M_{\rm BH} \sim 6.00$} \ \textsuperscript{\textit{c}}  \\ \hline

 \textbf{$\dagger$} 0780561301 & 125.29 & 0.9991 & 6.0 - 10.0 & $7.81\times10^{-5}$ & $9.77\times10^{-6}$ & 0.201 & 8.00 \\
 0780561301 & 125.29 & 0.9980 & 2.2 - 6.4 & $2.83\times10^{-4}$ & $9.77\times10^{-6}$ & 0.121 & 29.0 \\
 \textbf{$\dagger$} 0780561401 & 82.92 & 1.0000 & 4.0 - 5.8 & $7.81\times10^{-5}$ & $1.95\times10^{-5}$ & 0.279 & 4.00 \\
 0780561401 & 82.92 & 0.9997 & 3.2 - 9.8 & $4.69\times10^{-4}$ & $1.95\times10^{-5}$ & 0.136 & 24.1 \\
 0780561701 & 113.20 & 0.9995 & 3.2 - 3.8 & $9.77\times10^{-5}$ & $9.77\times10^{-6}$ & 0.209 & 10.0 \\
 \textbf{*} 0792180201 & 128.63 & 0.9979 & 4.0 - 5.6 & $1.66\times10^{-4}$ & $9.77\times10^{-6}$ & 0.294 & 17.0 \\
 \textbf{$\dagger$} 0792180301 & 53.26 & 0.9994 & 4.8 - 9.6 & $9.77\times10^{-5}$ & $1.95\times10^{-5}$ & 0.304 & 5.01 \\
 0792180501 & 93.83 & 0.9992 & 0.9 - 1.6 & $1.02\times10^{-3}$ & $1.95\times10^{-5}$ & 0.059 &  52.3 \\
 0792180601 & 105.02 & 0.9984 & 6.6 - 8.6 & $9.77\times10^{-5}$ & $9.77\times10^{-6}$ & 0.246 & 10.0 \\
 
 \hline
 \textbf{1H 0707-495, \ $M_{\rm BH} \sim 6.3$} \ \textsuperscript{\textit{c}}  \\ \hline

 0653510301 & 95.19 & 0.9984 & 4.8 - 6.2 & $1.95\times10^{-4}$ & $1.95\times10^{-5}$ & 0.255 & 10.0 \\
 0653510501 & 94.72 & 0.9997 & 4.0 - 5.4 & $1.95\times10^{-4}$ & $1.95\times10^{-5}$ & 0.225 & 10.0 \\
 \textbf{*} 0653510601 & 105.85 & 0.9975 & 5.6 - 8.2 & $1.27\times10^{-4}$ & $9.77\times10^{-6}$ & 0.224 & 13.0 \\

 \hline
 \textbf{PG1244+026, \ $M_{\rm BH} \sim 6.24$} \  \textsuperscript{\textit{d}}  \\ \hline

 0675320101 & 71.81 & 0.9993 & 3.0 - 5.8 & $2.15\times10^{-4}$ & $1.95\times10^{-5}$ & 0.074 & 11.0 \\

 \hline
 \textbf{NGC~4051, \ $M_{\rm BH} \sim 6.24$} \ \textsuperscript{\textit{e}}  \\ \hline

 0109141401 & 25.77 & 0.9997 & 5.0 - 6.0 & $3.52\times10^{-4}$ & $3.91\times10^{-5}$ & 0.095 & 9.00 \\
 \textbf{$\dagger$} 0606321801 & 39.83 & 0.9998 & 3.4 - 4.0 & $1.56\times10^{-4}$ & $3.91\times10^{-5}$ & 0.180 & 3.99 \\

 \hline
 \textbf{ARK~564, \ $M_{\rm BH} \sim 6.41$} \ \textsuperscript{\textit{f}}  \\ \hline

 \textbf{$\dagger$} 0006810101 & 4.95 & 1.0 & 3.0 - 5.4 & $1.88\times10^{-3}$ & $3.13\times10^{-4}$ & 0.134 & 6.00 \\
 0670130501 & 17.18 & 0.9987 & 2.0 - 3.8 & $8.59\times10^{-4}$ & $7.81\times10^{-5}$ & 0.073 & 11.0\\
 0670130701 & 44.71 & 1.0 & 2.6 - 2.8 & $2.73\times10^{-4}$ & $3.91\times10^{-5}$ & 0.163 & 6.98 \\
 0670130801 & 51.65 & 0.9998 & 3.0 - 3.4 & $2.54\times10^{-4}$ & $1.95\times10^{-5}$ & 0.072 & 13.0 \\
 
 \hline
 \textbf{MRK~766, \ $M_{\rm BH} \sim 6.63$} \ \textsuperscript{\textit{g}}  \\ \hline

 \textbf{*$\dagger$} 0304030301 & 38.35 & 0.9976 & 3.0 - 4.6 & $4.69\times10^{-4}$ & $3.91\times10^{-5}$ & 0.055 & 12.0 \\

 \hline
  
\end{tabular*}

\label{tab:QPOs}
\end{table*}

\section{Results}

From a total of 200 observations across our sample of 38 AGN, we detect 21 QPO candidates at a $\ge 3\sigma$ significance, and 3 at a borderline $3\sigma$ level, all from 7 AGN. We re-confirm the well-reported QPO in 4 observations of RE~J1034+396, with up to 20 additional detections across a further 6 AGN. Our initial results based on the procedure described above, are shown in Table \ref{tab:QPOs} which indicate the energy band in which the detection significance is maximised, the SMBH mass and the fractional rms in the QPO candidate (found by integrating the QPO candidate's power and removing the Poisson noise). Although we analyse all available observations of MS 2254.9-3712, our restriction of using only continuous data prevents us from locating the QPO reported in \citet{2015AlstonMS2254} as, after flare subtraction, we have a lightcurve of only 27.8~ks in comparison to the 63~ks combined EPIC-PN/EPIC-MOS segment defined in \citet{2015AlstonMS2254} where gaps were removed via interpolation.

We note that these results are only initial, and change in significance as we perform the additional tests in the case of each AGN, as we shall proceed to describe.

\subsection{RE~J1034+396}

\subsubsection{Initial {\sc plc} search}

To test our approach for locating QPO candidates, we examined the eight observations of RE~J1034+396, comparing our results to those of \citet{2014Alston5} noting once again that we do not include interpolation (which can restrict our frequency range). 

\begin{figure}
    \centering
    \includegraphics[width=0.47\textwidth]{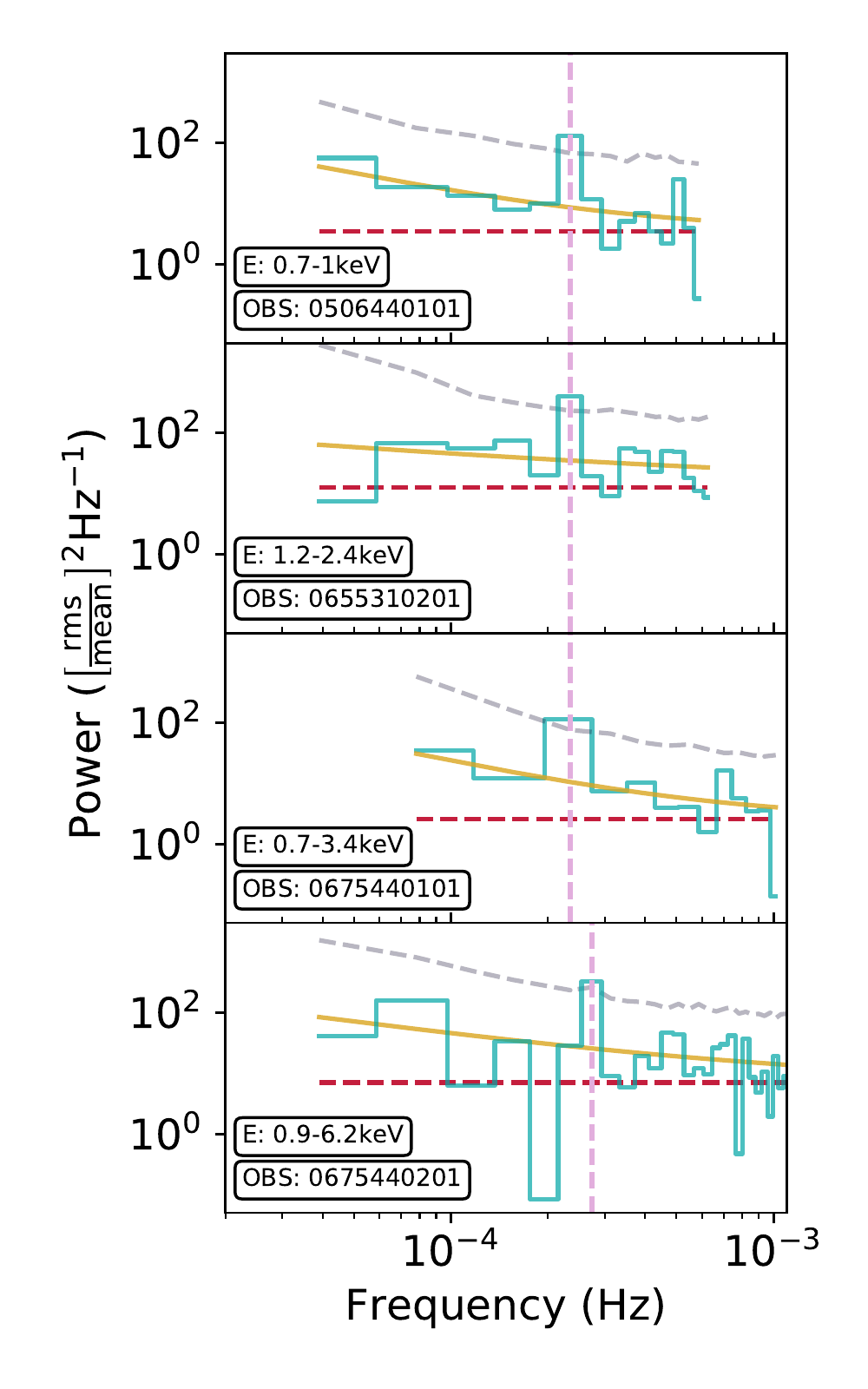}
    \caption{The highest significance ($\ge 3\sigma$) QPO candidates detected in RE~J1034+396, using the {\sc plc} model, with the corresponding frequency bin highlighted. The consistency of the QPO frequency across each observation has historically set RE~J1034+396 apart as it is highly improbable to create false peaks in power across multiple observations at the same frequency.}
    \label{fig:REJ_multiplot}
\end{figure}

We detect the previously reported QPO to a $>3\sigma$ confidence, at a frequency of 2-3$\times$10$^{-4}$ Hz in four observations -- see Figure \ref{fig:REJ_multiplot}. Notably, the maximum significance energy band extends down to 0.7 keV and reflects the presence of the QPO in the soft X-ray component \citep[see][]{2011Middleton8Years}. \citet{2014Alston5} detect a QPO in five of the eight observations (with the QPO now detected in new observations - \citealt{2020JinREJ}); including the four where we find a significant candidate. Whilst we do not detect a QPO in OBSID:0655310101, we select only the first $28.2$ks as a flare-less segment, and subsequently detect a QPO at only a $>2\sigma$ level.

\begin{figure*}
    \centering
    \includegraphics[width=138mm] {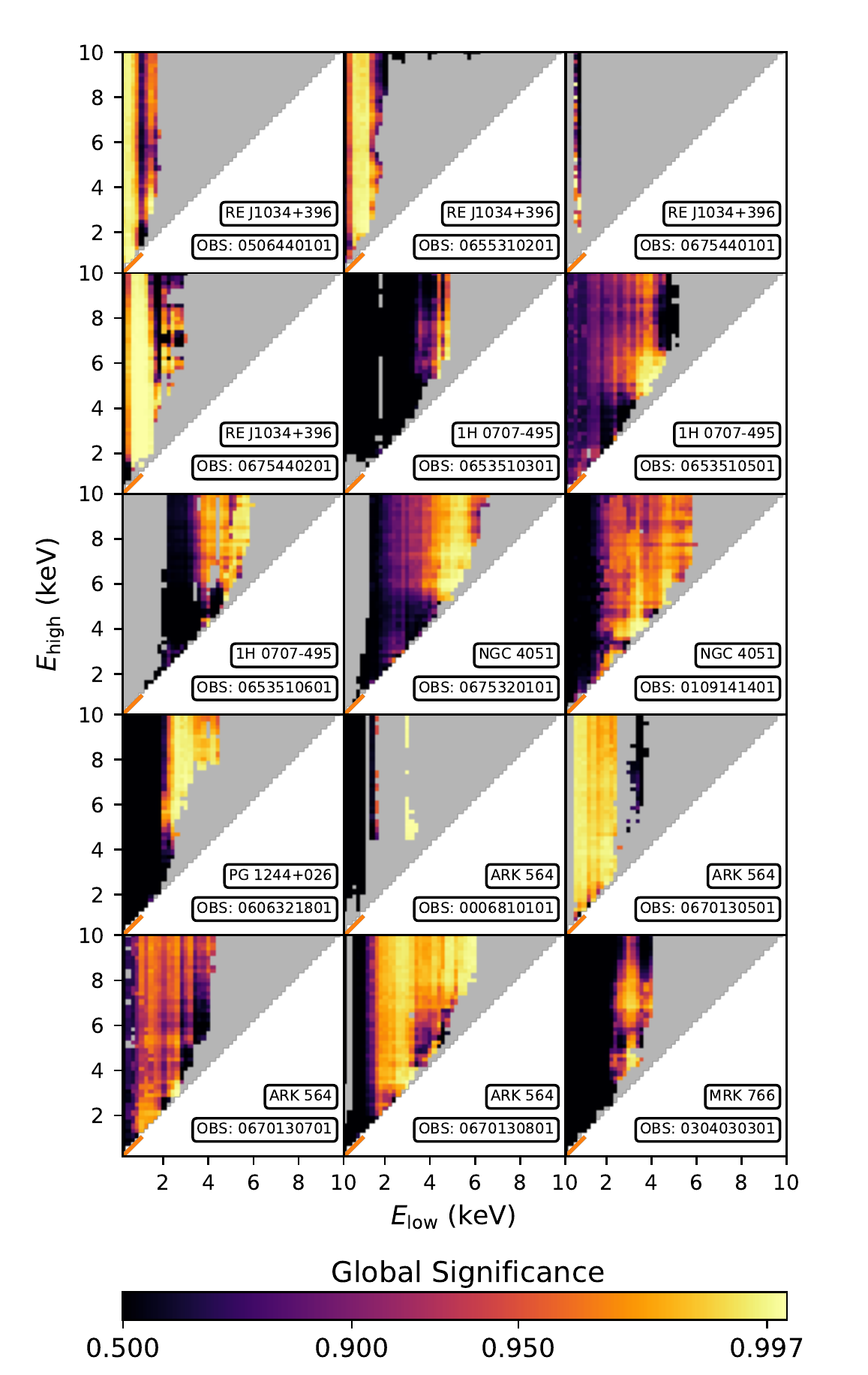}
    \caption{Energy-resolved significance heatmaps displaying the measured global significance of QPO candidates (excluding IRAS~13224-3809, for which see Figure \ref{fig:IRAS_Heatmap_panels}) using the {\sc plc} model, at the frequency of the highest significance blind detection in an energy bin defined as E$_{\rm low}~\rightarrow~$E$_{\rm high}$ The grey areas, including `gaps' in the heatmaps, indicate energy bands which do not fit the {\sc plc} criteria we set out in Section 3 -- including both white noise dominated observations and {\sc bknplc} model preferred fits.}
    \label{fig:Heatmap_panels}
\end{figure*}

In Figure \ref{fig:Heatmap_panels}, we present energy-resolved significance heat-maps which indicate the $\ge 3\sigma$ significant detections detailed in Table \ref{tab:QPOs}, as well as the global significance of any signal at this specific frequency in all the other energy bands. We note that the significances we present in Figure \ref{fig:Heatmap_panels} are purely blind (i.e. the full number of independent trials is still taken into account). 

\subsubsection{Further tests}

As the QPO in RE~J1034+396 is well established and thoroughly investigated in the literature (e.g. \citealt{2010BayesianVaughan}), we refrain from applying our additional tests in this instance.

\bigskip
\noindent {\bf Frequency-restricted search:} Given our detection of multiple $\ge 3\sigma$ QPO candidates at consistent frequencies, we make the reasonable assumption that we have \textit{a-priori} knowledge of the QPO frequency in RE~J1034+396, across any observation. We then test whether a signal is located at this frequency -- in this case the modal frequency of $2.34\times$10$^{-4}$ Hz (or the closest Fourier frequency) -- in those observations without blind detections. In effect, this separate p-value test lowers the effective threshold for detection, as it reduces the number of free trials to a single Fourier bin of interest. However, this further analysis does not locate any additional QPO candidates at $\ge 3\sigma$.

\subsection{IRAS 13224-3809}

IRAS~13224-3809 is one of the most variable NLS1 AGN and has been the focus of a recent 1.5~Ms study with \textit{XMM Newton}, yielding high quality data with almost full orbits (see \citealt{2019AlstonIRASvariability}, \citealt{2020AlstonIRASreverbmap}). We studied 16 available observations, and in 6 of these detect QPO candidates at $\ge 3\sigma$, with one additional observation with a borderline $3\sigma$ detection.

\subsubsection{Initial {\sc plc} search}

\begin{figure}
    \centering
    \includegraphics[width=0.45\textwidth, trim={0cm 0.5cm 0cm 0}]{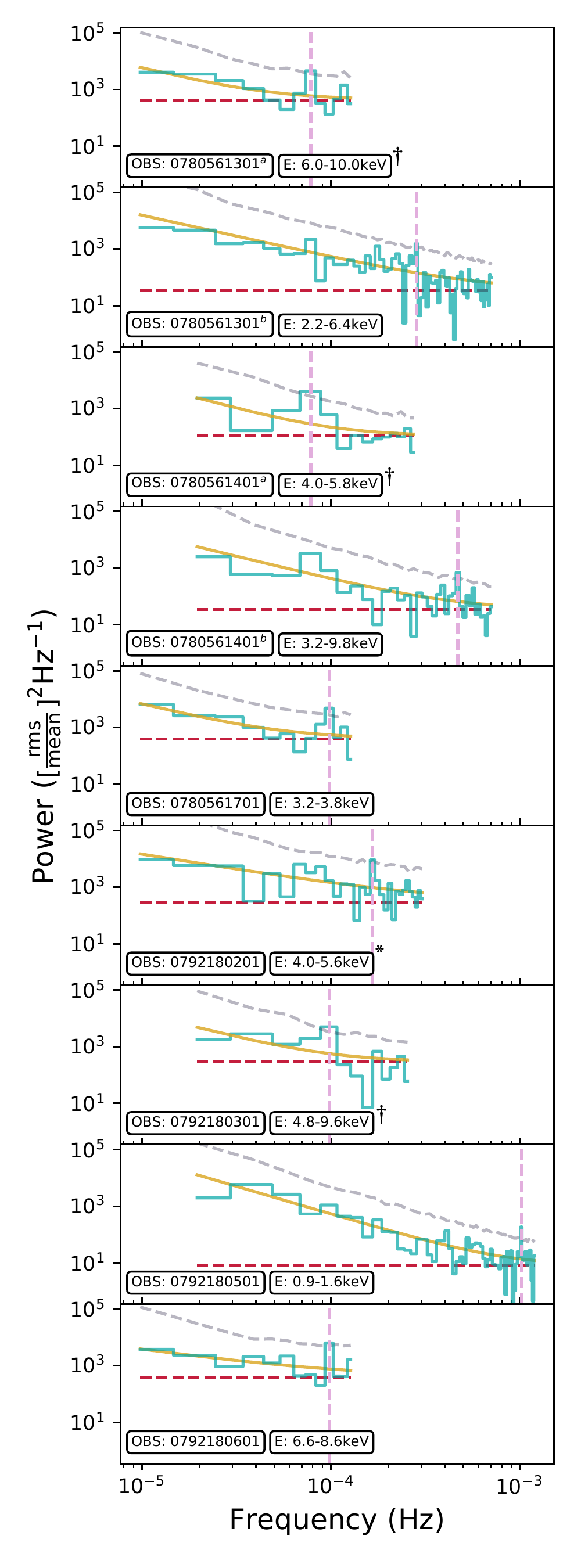}
    \caption{The highest significance ($\ge 3\sigma$) QPO candidates found in IRAS 13224-3809 using the {\sc plc} model, with the corresponding frequency bin highlighted. We note the consistency of frequencies across multiple observations, with the clear exception of OBSID:0792180501. See Table \ref{tab:QPOs} for supplementary information. We highlight borderline $3\sigma$ results (\textbf{*}) and those which are not validated by our further tests (\textbf{$\dagger$} -- see Section 3.2). }
    \label{fig:IRAS_multiplot}
\end{figure}

Three observations all have consistent QPO candidate frequencies at $9.77\times$10$^{-5}$ Hz (OBSIDs:0780561701, 0792180301 and 0792180601) with an extremely high fractional rms ($>$ 20\%) -- see Figure \ref{fig:IRAS_multiplot} In addition to these, we find a borderline $3\sigma$ QPO candidate with a frequency of $1.66\times$10$^{-4}$ Hz in OBSID:0792180201; although the significance fluctuates around $3\sigma$, it sits close to the frequency of QPO candidates detected in other observations and is therefore of interest.

\begin{figure*}
    \centering
    \includegraphics[width=0.3\textwidth,trim={6cm 0 6cm 0}]{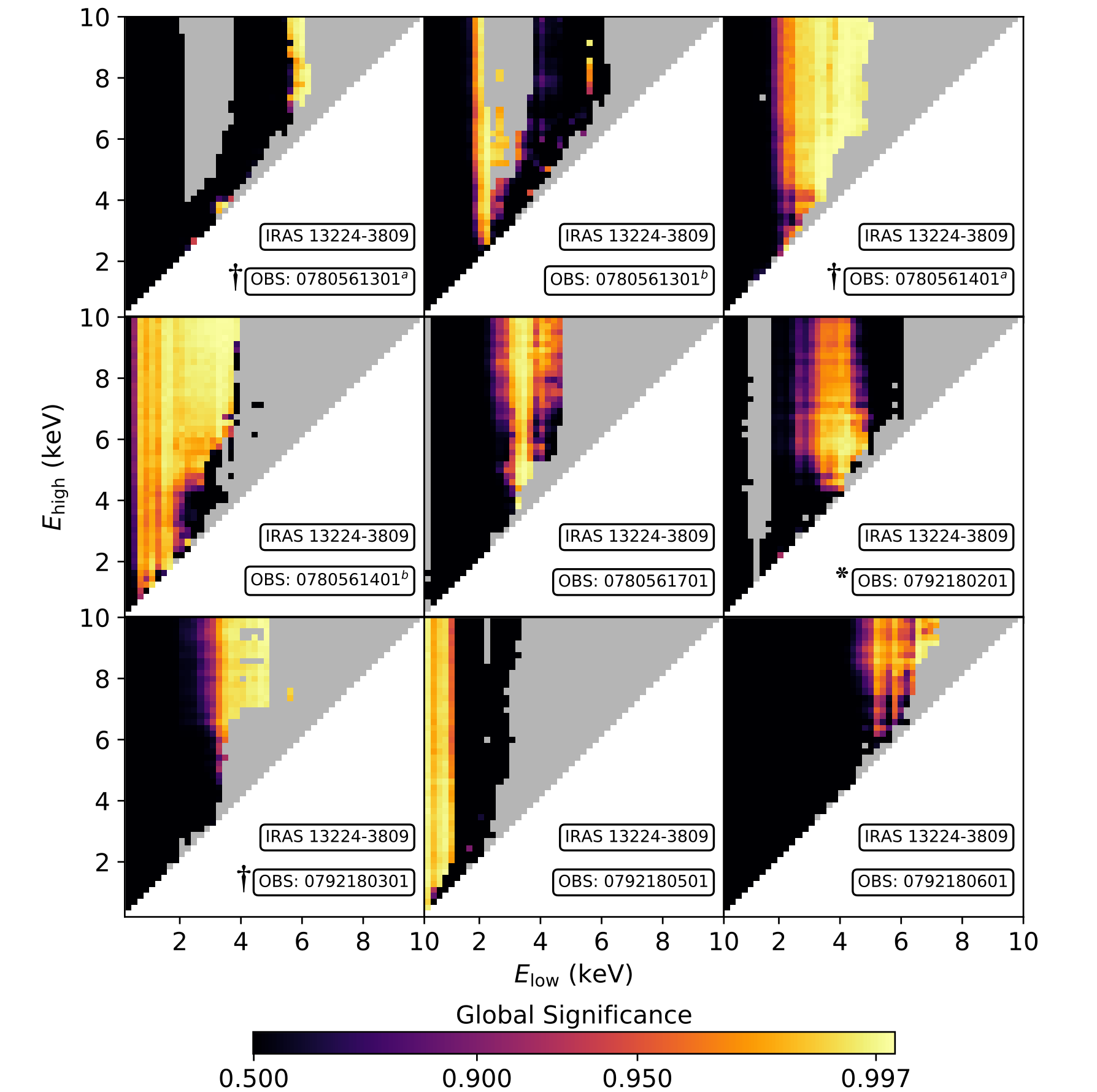}
    \caption{Energy-resolved significance heatmaps displaying the measured global significance of the QPO candidates in IRAS 13224-3809, using the {\sc plc} model, at the frequency of the highest significance blind detection, in an energy bin defined as E$_{\rm low}~\rightarrow~$E$_{\rm high}$. The grey areas, including `gaps' in the heatmaps, indicate energy bands which do not fit the {\sc plc} criteria we set out in Section 3 -- including both white noise dominated observations and {\sc bknplc} model preferred fits.}
    \label{fig:IRAS_Heatmap_panels}
\end{figure*}

We also detect a $> 3\sigma$ QPO candidate at $7.81\times$10$^{-5}$ Hz in two observations (OBSID:0780561301 and 0780561401). These two observations provide a potentially intriguing result in that they contain $>3\sigma$ QPO candidates at {\it two} different frequencies, dependent on the energy band selected, both of which are presented in Table \ref{tab:QPOs}. In OBSID:0780561301, the higher frequency QPO candidate is found at $2.83\times$10$^{-4}$ Hz, whilst in OBSID:0780561401 it is found at $4.69\times$10$^{-4}$ Hz. If we assume these to be harmonically related to the $7.81\times$10$^{-5}$ Hz feature then these would be at 7:2 and 6:1 resonances respectively. Figure \ref{fig:IRAS_Heatmap_panels} displays the energy dependence of the global significance of these QPO candidates; the energies at which they become statistically significant at $\ge3\sigma$, are almost entirely confined to hard energies $>3$~keV. The notable exception to the energy-dependence is that of OBSID:0792180501, which appears to show several features that are unique amongst our sample.

\begin{figure*}
    \centering
    \includegraphics[width=\textwidth]{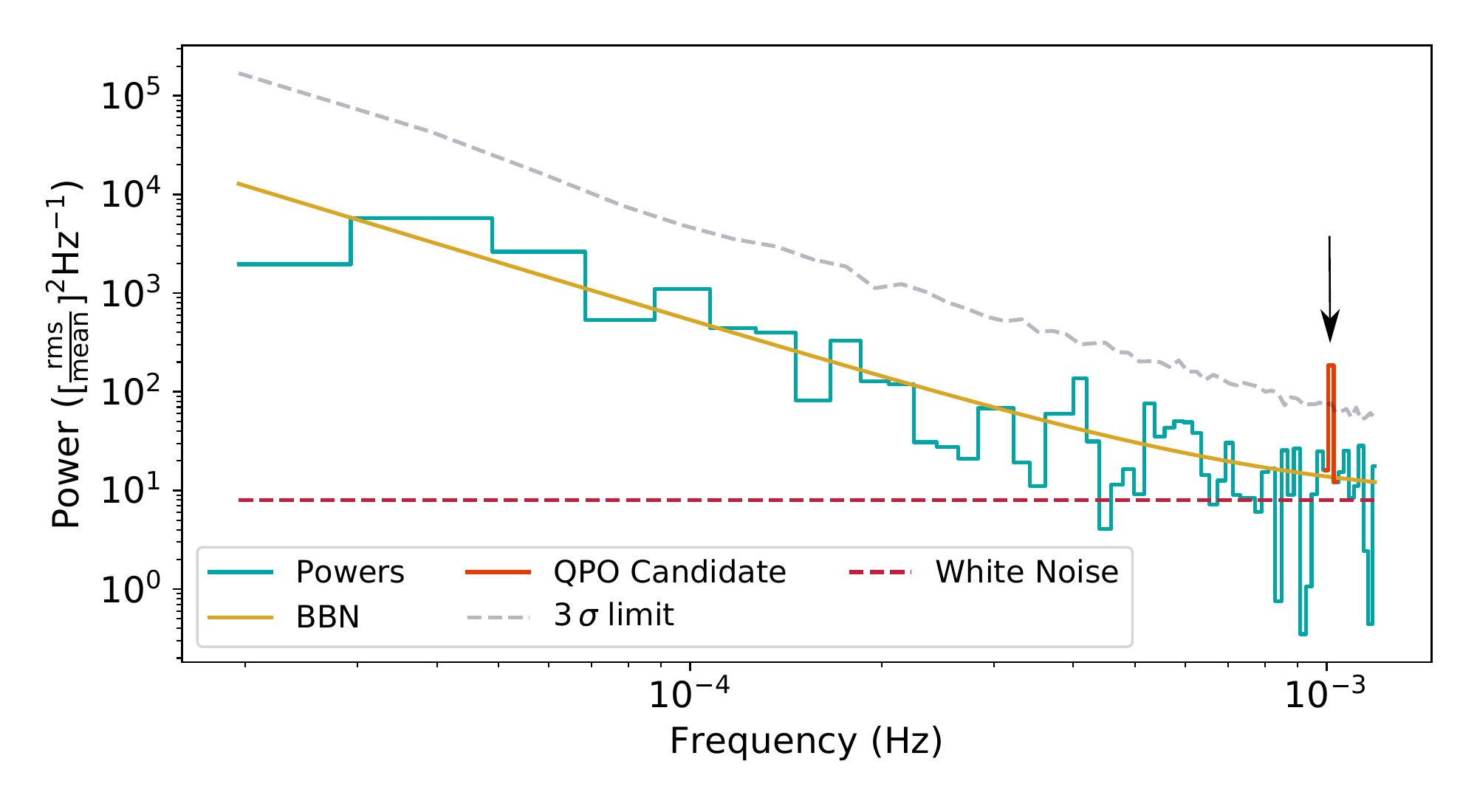}
    \caption{The highest significance QPO candidate (in the $0.9 - 1.6$ keV band) found using the {\sc plc} model, in OBSID:0792180501 of IRAS 13224-3809. This standout candidate is observed with Q $>$ 50 and is one of only two examples we have found of a QPO candidate at $> 1\times$10$^{-3}$ Hz. See Table \ref{tab:QPOs} for supplementary information.}
    \label{fig:HFQPO}
\end{figure*}

The QPO candidate in OBSID:0792180501 is found at a still higher frequency of $1.02\times$10$^{-3}$ Hz, as shown in Figure \ref{fig:HFQPO}. This QPO candidate is almost unique amongst our detections (the other example being found in ARK 564), as it lies at a frequency higher than the Poisson noise cutoff for the vast majority of our observations, and we only detect it due to the high count rates. In combination with the length of this observation (setting the size of the natural Fourier frequency bin), this QPO candidate has the largest $Q$ value of any of our detections ($Q > 50$). The energy band at which the significance of this QPO candidate is maximised also differs from those of the lower frequency QPO candidates, being found at softer energies (specifically $0.9 - 1.6$ keV -- see Figure \ref{fig:IRAS_Heatmap_panels}) without significant detections above 2~keV (although at these higher energies, the data quality becomes insufficient to study such high frequencies). It is of interest to note that those QPO candidates detected at lower frequencies do not appear in this observation.

\subsubsection{Further tests}

We proceed to apply the two additional tests discussed in Section 3.2 and describe the results on an observation-by-observation basis.

\medskip
\noindent{\bf Test(i)}: 

\begin{itemize}

\item OBSID:0780561301. We find a break at  $\sim2\times$10$^{-4}$ Hz at energies $\sim 2-6$~keV. In this case we observe a sudden drop in power above the high frequency QPO candidate at $2.83\times$10$^{-4}$ Hz, as shown in the lower panel of Figure \ref{fig:IRAS_Obs6}. This behaviour may not be best described by a break but is similar to the findings of \citet{2012GonzMartinVaughan} when modelling the PSD of MS 22549-3712, in which they report a poorly constrained fit using a bending-power law with an extremely steep index above the break, but no QPO. Assuming a free \textsc{bknplc} model does not lower the significance of the higher frequency candidate QPO in this observation. In the case of energies $>$6~keV, we find a break at $\sim 2\times10^{-5}$ Hz or $\sim 6\times10^{-5}$ Hz, and in doing so, the significance of the QPO candidate at $7.81\times10^{-5}$ Hz drops to below $3\sigma$ -- as shown in upper panel of Figure \ref{fig:IRAS_Obs6}. By including a break, the white noise cutoff (i.e. the frequency at which the Poisson noise dominates over the red noise) is brought to lower frequencies and restricts the number of frequency bins in the periodogram; as a result, we find multiple energy bands which contained $\ge 3\sigma$ QPO candidates using the \textsc{plc} model, to now be white noise dominated, i.e. with $<$ 13 Fourier bins below the cutoff (which technically includes the QPO candidate bin, although the power in this bin is ignored in defining the null hypothesis). 

\begin{figure}
    \centering
    \includegraphics[width=0.47\textwidth]{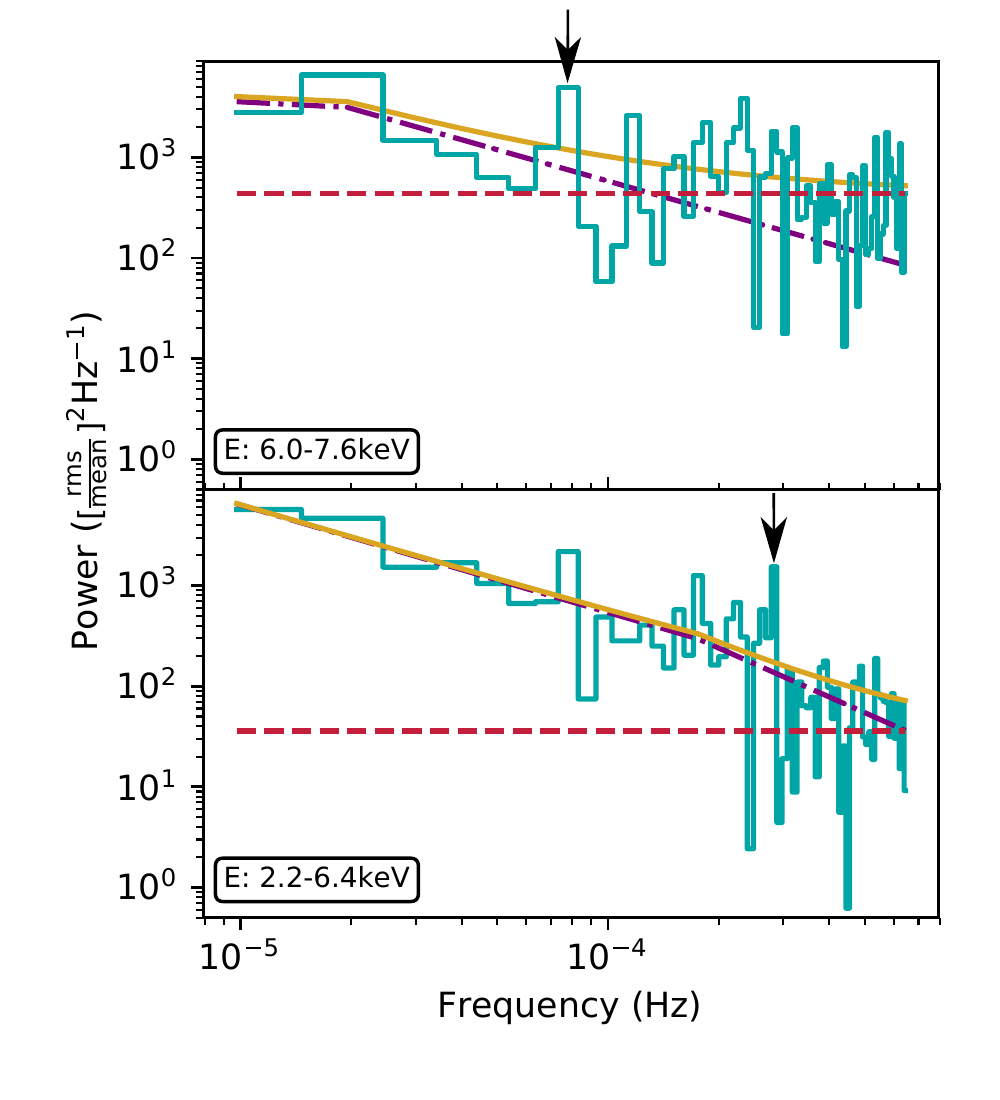}
    \caption{Energy resolved periodograms of IRAS 13224-3809 OBSID:0780561301, fitted with \textsc{bknplc} models following test (i). Top: periodogram in the 6.0 - 7.0~keV band, with the QPO candidate highlighted at $7.81\times10^{-5}$ Hz. We find fitting with this model  causes the QPO candidate to drop to a $2\sigma$ significance. Bottom: periodogram in the 2.2 - 6.4~keV band, with the QPO candidate highlighted at $2.83\times10^{-4}$ ~Hz, which remains $> 3\sigma$ before a steep drop in power at higher frequencies. Note: this frequency lies within the white noise in the higher energy band periodogram (top panel).}
    \label{fig:IRAS_Obs6}
\end{figure}

\item OBSID:0780561401. We find the high-frequency QPO candidate at $4.69\times$10$^{-4}$ Hz lies consistently above the $3\sigma$ threshold when within the white noise limit, see upper panel of Figure \ref{fig:IRAS_Obs7}. For energy ranges above 3.2~keV however, we find the white noise cutoff moves to frequencies below $4.69\times$10$^{-4}$ Hz and the QPO candidate is no longer considered. We also find that the low-frequency QPO candidate at $7.81\times10^{-5}$ Hz may be well described by a break across an energy range of $2 - 10$~keV, although the strength of the QPO candidate increases with increasing energies and becomes less well described by a break given our methodology, see lower panel of Figure \ref{fig:IRAS_Obs7}. Given the result of this test, we remain cautious of the validity of this low-frequency QPO candidate and believe it requires further investigation. 

\begin{figure}
    \centering
    \includegraphics[width=0.47\textwidth]{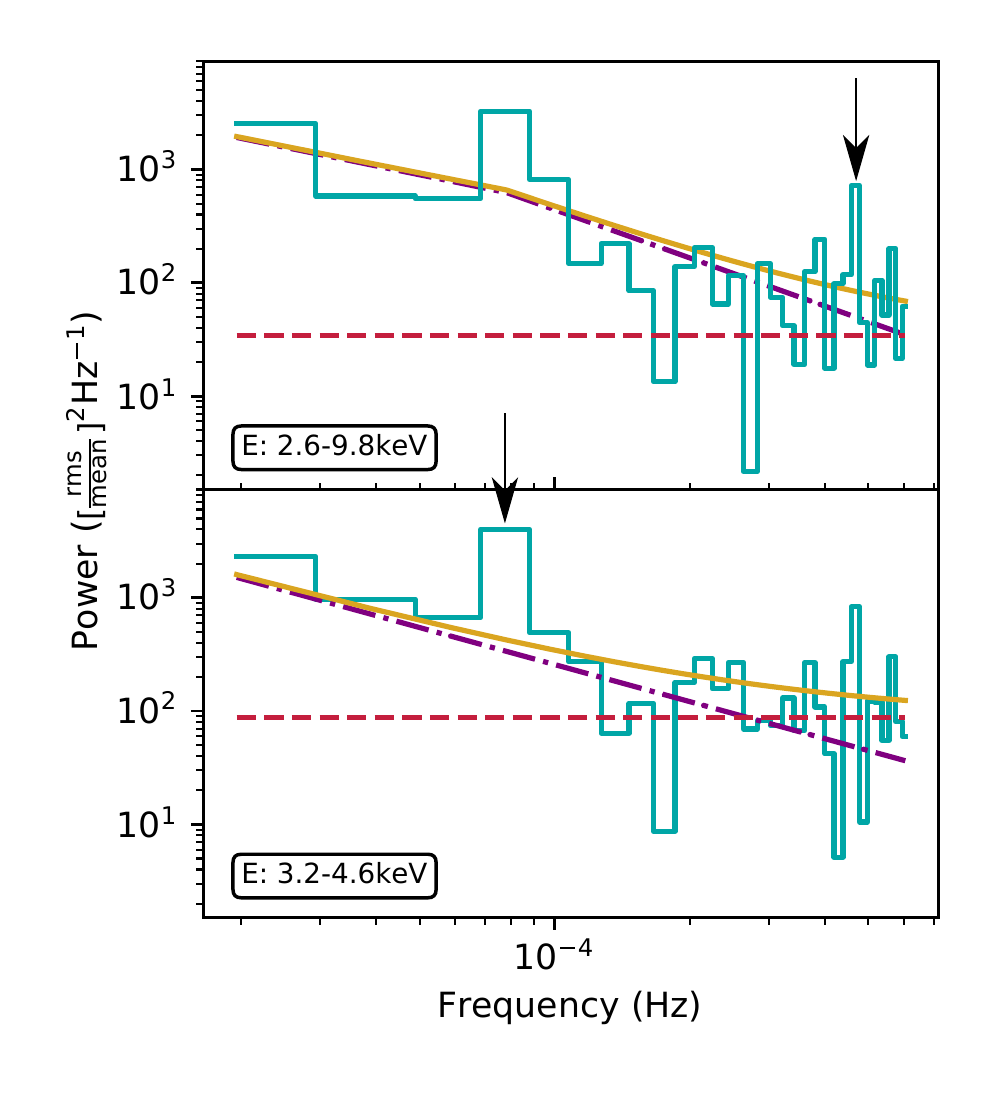}
    \caption{Energy resolved periodograms of IRAS 13224-3809 OBSID:0780561401, fitted with \textsc{bknplc} models following test (i). Top: periodogram in the 2.6 - 9.8~keV band, with the QPO candidate highlighted at $4.69\times10^{-4}$~Hz, found at a $> 3\sigma$ significance using a \textsc{bknplc} model, with the break located at $7.81\times10^{-5}$~Hz. Bottom: periodogram in the 3.2 - 4.6~keV band; the QPO candidate is now found at the position of the break (contrast with the top panel) at $7.81\times10^{-5}$~Hz. In this energy band, a break does not well describe the data and, following test (i), we find the QPO candidate at $7.81\times10^{-5}$~Hz reaches a $> 3\sigma$ significance. Note: the previous QPO candidate at $4.69\times10^{-4}$ Hz is beyond the white noise cutoff in this energy band (contrast with the top panel).}
    \label{fig:IRAS_Obs7}
\end{figure}

\item OBSID:0780561701. We find a weak break (i.e. a  value of $\beta_{1}/\beta_{2}$ close to unity) at $\sim 2\times10^{-5}$ Hz and at energies $>$ 3~keV, which has no impact on the significance of the QPO candidate at $9.77\times10^{-5}$ Hz. 

\item OBSID:0792180201. We reported only a borderline detection using a \textsc{plc} model. We subsequently find a break at $\sim8\times$10$^{-5}$ Hz, however, the QPO candidate at $1.66\times10^{-4}$ Hz becomes {\it more} significant. 

\item OBSID:0792180301. We lose the QPO candidate at $9.77\times10^{-5}$ Hz entirely with a weak break at $\sim 4\times10^{-5}$ Hz and before a steep drop in power at frequencies above the QPO candidate. We find that neither the \textsc{plc} or \textsc{bknplc} models provide a good description of the broad band noise, but in assuming a \textsc{bknplc} model, the white noise cutoff is brought to sufficiently low frequencies for the data to be white noise dominated.

\item OBSID:0792180501. We find a weak break around $\sim 4\times10^{-5}$ Hz, but the QPO candidate at $1.02\times10^{-3}$ Hz remains significant. 

\item OBSID:0792180601. The lower frequency white noise cutoff means we are close to datasets becoming white noise dominated, however, the QPO candidate at $9.77\times10^{-5}$ Hz remains significant on the edge of our frequency range. 
\end{itemize}

\bigskip
\noindent{\bf Test(ii):} 

All of the results with a fixed $\beta_{1}$ agree with our findings from test(i), (as expected given $\beta_{1}$ = -1.1 was found previously in the case of IRAS 13224-3809 \citealt{2019AlstonIRASvariability}) although this assumed shape for the broad band noise does not always provide a good description depending on the energy range under consideration.

\bigskip
The QPO candidates identified through our initial search using the {\sc plc} model, but where the above tests question their validity, are highlighted in Table \ref{tab:QPOs} and the respective figures.

\bigskip
\noindent{\bf Frequency-restricted search:} As with RE~J1034+396, the detection of multiple $\ge 3\sigma$ QPO candidates at consistent frequencies, allows us to assume \textit{a-priori} knowledge of a possible QPO candidate at frequencies of $7.81\times$10$^{-5}$~Hz and $9.77\times$10$^{-5}$ Hz (or the closest Fourier frequency). Assuming a frequency of $9.77\times$10$^{-5}$~Hz, we find a frequency-restricted $\ge3 \sigma$ QPO candidate at energy ranges $> 3$~keV in OBSID:0673580401. We note that this observation does not show any $\ge 3\sigma$ global significance features, so does not appear in Table \ref{tab:QPOs}. Assuming a frequency of $7.81\times$10$^{-5}$ Hz, we find a single frequency-restricted $\ge3 \sigma$ QPO candidate at 2.8 - 3~keV in OBSID:0792180201. 

In light of the possible harmonic features observed in OBSIDs:0780561301 and 0780561401, we also search for QPO candidates at frequencies of $2.83\times$10$^{-4}$ Hz and $4.69\times$10$^{-4}$ Hz respectively. At a frequency of $2.83\times$10$^{-4}$~Hz, we find frequency-restricted $\ge 3\sigma$ candidates in OBSIDs: 0673580101 and 0780561601, both in the 3 - 6~keV range, comparable to the QPO candidate in OBSID:0780561301. At a frequency of $4.69\times$10$^{-4}$ Hz, we also find frequency-restricted $\ge 3\sigma$ candidates in OBSID:0673580101 in the 2 - 8~keV range, and in OBSID:0780560101, at energies around 2~keV. The apparent detection of a higher frequency QPO candidate in the {\it same} observation (OBSID:0673580101) but at {\it both} frequencies highlights the potential problem with searching at a reduced number of frequencies: even with an {\it a priori} expectation -- the significance is artificially higher and care must be taken when claiming any detection (i.e. independent evidence is still required).

\subsection{1H 0707-495}

1H~0707-495 has been studied extensively by \textit{XMM-Newton} with a total observing time in excess of 1~Ms (e.g. \citealt{Kara20131H1Msdata}).

\subsubsection{Initial {\sc plc} search}

\begin{figure}
    \centering
    \includegraphics[width=0.47\textwidth]{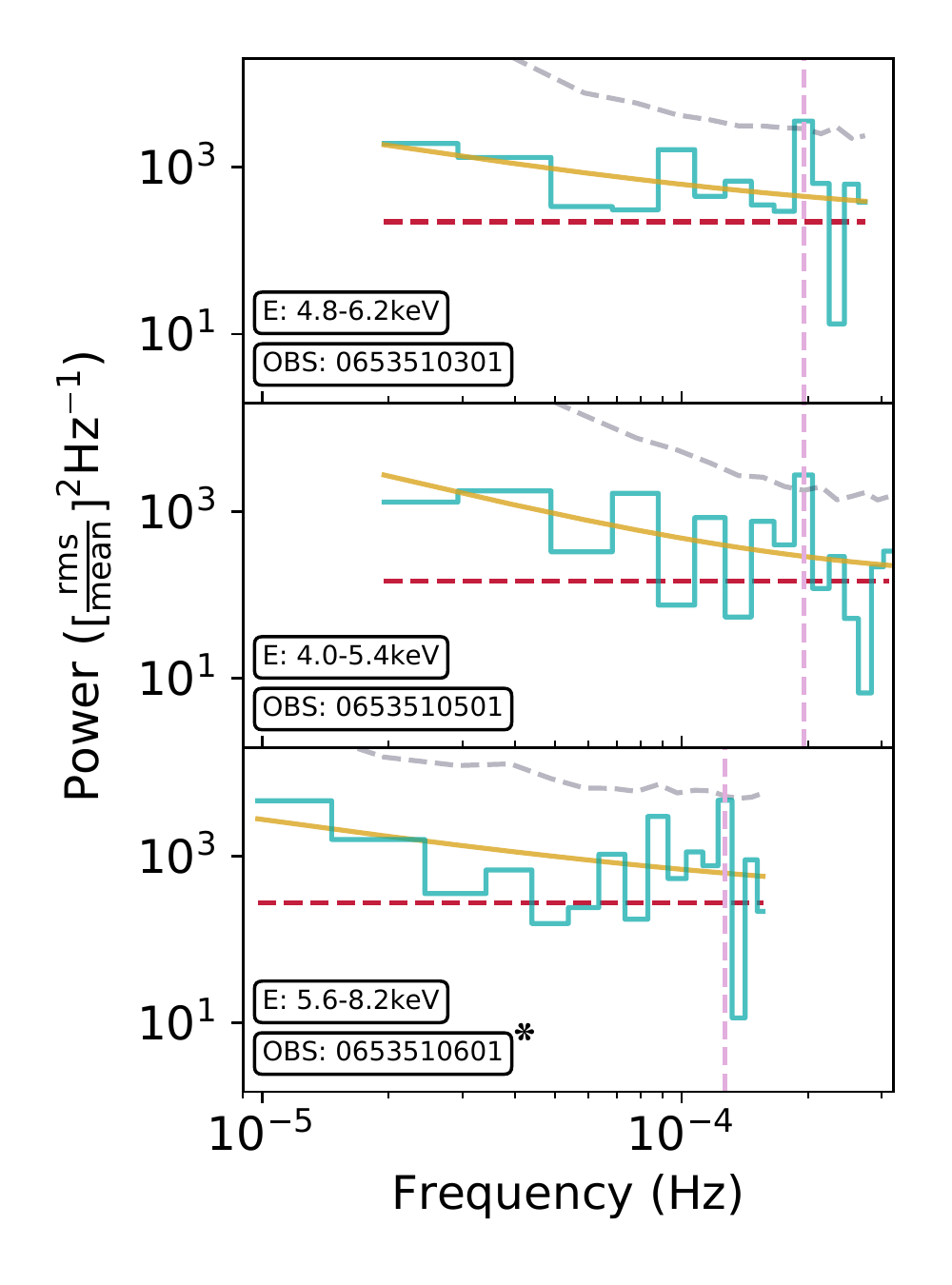}
    \caption{The highest significance ($\ge 3\sigma$) QPO candidates detected in 1H~0707-495 using the {\sc plc} model, with the corresponding frequency bin highlighted. As with RE~J1034+396 and IRAS 13224-3809, we note the consistency of the QPO candidate frequency across multiple observations. We highlight borderline $3\sigma$ results (\textbf{*}). }
    \label{fig:1H_multiplot}
\end{figure}

As shown in Figure \ref{fig:1H_multiplot}, we detect $\ge3\sigma$ QPO candidates in three observations (including one borderline detection), from a total of 11, with high fractional rms ($>$ 20\%) and across a narrow frequency range of 1-2$\times$10$^{-4}$ Hz. 

We find a borderline $\sim 3\sigma$ QPO candidate in OBSID:0653510601, although in this case, the QPO candidate frequency lies close to or even at the white noise cutoff in some energy bins. For this observation, we find a slight shift towards lower frequencies, but there is minimal change in the fractional-rms of the QPO candidate between each of these observations. 

We find the peak QPO candidate significance in OBSID:0653510301 and 0653510501 to lie in the $\sim$4 - 6~keV range, with no $\ge 3\sigma$ energy bands extending below $3.8$\,keV, whilst we find only borderline $3\sigma$ features in OBSID:0653510601 in the $\sim$4 - 8~keV energy range. The extent of the significance variation is highlighted in Figure \ref{fig:Heatmap_panels}. We note also that the fractional-rms of the QPO candidates in 1H~0707-495 is similar to that found in IRAS~13224-3809.

\subsubsection{Further tests}

\medskip
\noindent{\bf Test(i)}:

\begin{itemize}
    \item OBSID:0653510301. Assuming a free \textsc{bknplc} model, we find a weak break at $\sim1\times$10$^{-4}$ Hz. The biggest impact is the shift of the white noise cutoff; in the 4.4 - 5.4~keV energy range the QPO candidate remains above the $3\sigma$ threshold but at the limit of non white noise dominated bins. As we move to harder energy bands, (namely 4.8 - 6.2~keV), the QPO candidate sits at the location of the white noise cutoff and we thus cannot discern its significance. 

\item OBSID:0653510501. In the 3.8 - 5.4~keV band, the periodogram is best described by a break at $\sim8\times$10$^{-5}$ Hz, whilst at 4.2 - 5.8~keV, the break extends back to $\sim3\times$10$^{-5}$ Hz. In both cases, the QPO candidate at $1.95\times$10$^{-4}$ Hz remains $\ge 3\sigma$. As with the previous observation, there are also energy bands in which we are now restricted by the white noise cutoff. This is seen in narrower energy bands (e.g. 3.8 - 4.8~keV and 4.2 - 5.4~keV), as the number of counts is restricted. In these cases, the QPO candidate remains $\ge 3\sigma$ but at a frequency closer to the white noise cutoff.

\item OBSID:0653510601. A borderline detection with a \textsc{plc} model, when assuming a \textsc{bknplc} model we find a tentative break around the QPO candidate frequency $\sim 1\times$10$^{-4}$ Hz. For this observation, we are investigating only two $\ge3 \sigma$ energy bands, where the QPO candidate is found at frequencies below the \textsc{plc} white noise cutoff. For the peak significance band (5.6 - 8.2~keV), we find the QPO candidate remains  $\ge$ 3$\sigma$, whilst in the 4.8 - 5.4~keV band, the QPO candidate falls below our detection threshold due to the presence of the break.

\end{itemize}

\bigskip
\noindent{\bf Test(ii)}:

\begin{itemize}
    \item OBSID:0653510301. Similar to the free-index case above, we find the QPO candidate remains significant above the broad band noise, but with a white noise cutoff located at a lower  frequency which reduces the number of frequency bins below our limit. 

\item OBSID:0653510501. Due to the forced upper index $\beta_1 = -1.1$ providing a poor description of the power spectrum of 1H~0707-495, we find only a limited number of breaks with this model. Notably, we find that the $\ge 3\sigma$ detection remains in the peak significance energy band (4.0 - 5.4~keV), but in others the index above the break is forced to be so steep as to bring the white noise cutoff down such that it would class as 'white noise dominated' by our definition.

\item OBSID:0653510601. Similar to OBSID:0653510301, we find the detection in the peak significance energy band (5.6 - 8.2~keV), remains above the $3\sigma$ threshold, but the periodogram is at the limit of being white noise dominated. 

\end{itemize}

\bigskip
\noindent{\bf Frequency-restricted search:} Based on our QPO candidate detections, we proceed to search the other observations of 1H~0707-495 without blind detections at the modal QPO candidate frequency of $1.95\times$10$^{-4}$ Hz. We detect a broad peak of power spanning this frequency in OBSID:0506200501 in the energy range of 4 - 9~keV, with a frequency-restricted p-value $> 3\sigma$ and an outlier with a frequency-restricted $> 3\sigma$ p-value in OBSID:0653510601 at this same frequency. The latter is of particular interest, as in OBSID:0653510601 we observe a borderline $3\sigma$ global significance detection at $1.27\times$10$^{-4}$ Hz, which is strongest in the $5.6 - 8.2$~keV energy range, but here we find the strongest frequency-restricted candidate at $3.4 - 4.0$~keV, which extends to higher energies with reduced significance. It appears this feature may be similar to those QPO candidate detections in OBSIDs:0653510301 and 0653510501, but is intrinsically weaker and becomes dominated by a candidate at $1.27\times$10$^{-4}$ Hz at higher energies.

\bigskip
\noindent{\bf Previous claims:} We note previous claims of significant QPO detections in 1H~0707-495, both concerning a single observation, OBSID:0511580401. Both \citet{2016Pan1HQPO} and \citet{Zhang20181H} report a QPO at $2.6\times10^{-4}$ Hz in the full 0.2 - 10~keV \textit{XMM-Newton} energy range, via a Fourier and a non-Fourier wavelet analysis respectively. We do not recover a $3\sigma$ result at this frequency, instead finding a $2\sigma$ outlier at energies $> 2.2$\,keV. We note that our blind selection criteria leads us to explore an 88.9~ks flare-subtracted segment in this observation, whereas \citet{2016Pan1HQPO} opt instead to focus on the first $\sim55$~ks. This selection maximises their QPO significance, whilst considering the full observation reduces it to $90\%$, well below the $3\sigma$ detection threshold. We will revisit the impact of such segment selection on QPO detection in a forthcoming paper.

\subsection{PG~1244+026}

\subsubsection{Initial {\sc plc} search}

\begin{figure}
    \centering
    \includegraphics[width=0.47\textwidth]{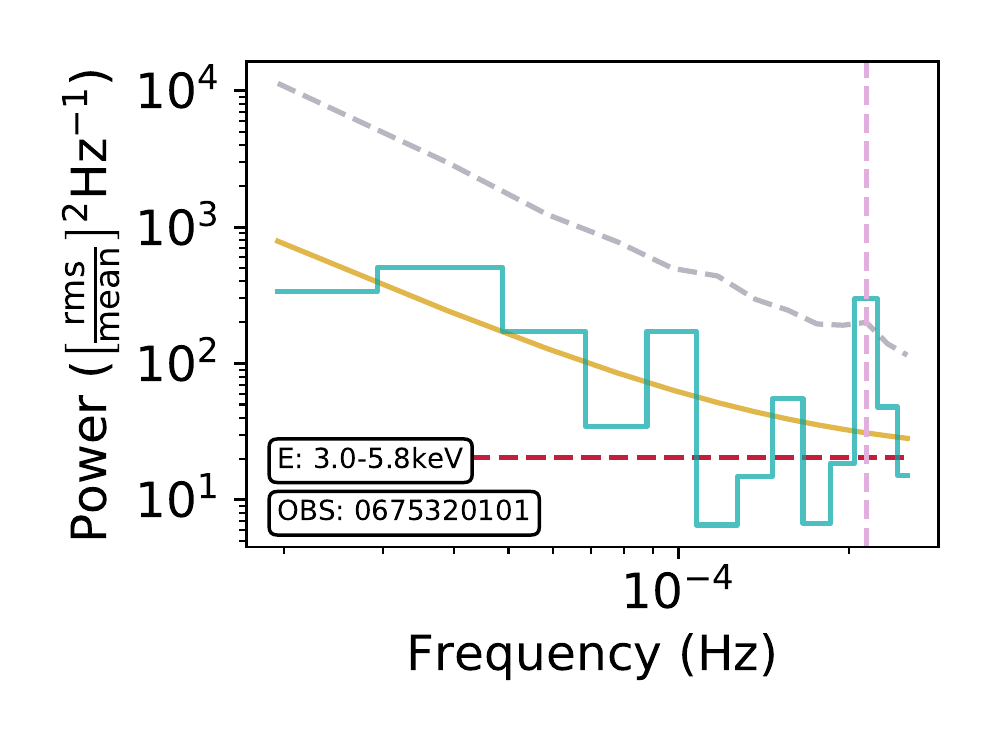}
    \caption{The highest significance ($\ge 3\sigma$) QPO candidate detected in a single observation of PG 1244+026 using the {\sc plc} model, with the corresponding frequency highlighted. }
    \label{fig:PG_singleplot}
\end{figure}

From a total of 7 \textit{XMM-Newton} observations of the NLS1 PG~1244+026, we detect only a single $\ge 3\sigma$ QPO candidate, within the 71\,ks longest continuous segment of OBSID:0675320101. We detect a QPO candidate at a frequency of $2.15\times10^{-4}$ Hz (see Figure \ref{fig:PG_singleplot}), similar to the frequencies of those found in RE~J1034+396, IRAS~13224-3809 and 1H~0707-495. The $\ge 3\sigma$ detections are largely restricted to energies above $\sim3\,$keV, as shown in Figure \ref{fig:Heatmap_panels}. This differentiates it from RE~J1034+396, IRAS~13224-3809 and 1H~0707-495, which also show $\ge 3\sigma$ detections at softer energies. 

\subsubsection{Further tests}

\medskip
\noindent{\bf Test(i)}:

\begin{itemize}
    \item OBSID:0675320101. In all cases where the QPO candidate is determined to be $\ge 3\sigma$, a weak break is found at $\sim4\times10^{-5}$ Hz. The QPO candidate remains highly significant in the $\sim2.0 - 7.0$~keV energy range, with reduced significance towards higher energies, e.g. $\sim2.0 - 9.0$~keV. However, in multiple energy bands, the QPO candidate now falls above the white noise cutoff or the number of frequencies becomes too small to avoid being white noise dominated by our definition. 

\end{itemize}

\bigskip
\noindent{\bf Test(ii)}:

\begin{itemize}
    \item OBSID:0675320101. We find the power spectrum may be well described with a lower index of -1.1 and the results from this test match those of test (i). 

\end{itemize}

\bigskip
\noindent{\bf Frequency-restricted search:} We proceed to set an \textit{a-priori} QPO candidate frequency (this time at $2.15\times10^{-4}$ Hz), and repeated the analysis for all observations of PG~1244+026. We find a single frequency-restricted $\ge 3\sigma$ candidate at this frequency in OBSID:0744440301, strongest in the 2.0 - 7.8~keV band.

\subsection{NGC 4051}

As with the cases of IRAS~13224-3809 and 1H~0707-495, NGC~4051 is one of best-studied, highly variable NLS1 known (e.g. \citealt{2010BreedtNGC4051}, \citealt{2011VaughanNGC4051}). 

\subsubsection{Initial {\sc plc} search}

\begin{figure}
    \centering
    \includegraphics[width=0.47\textwidth]{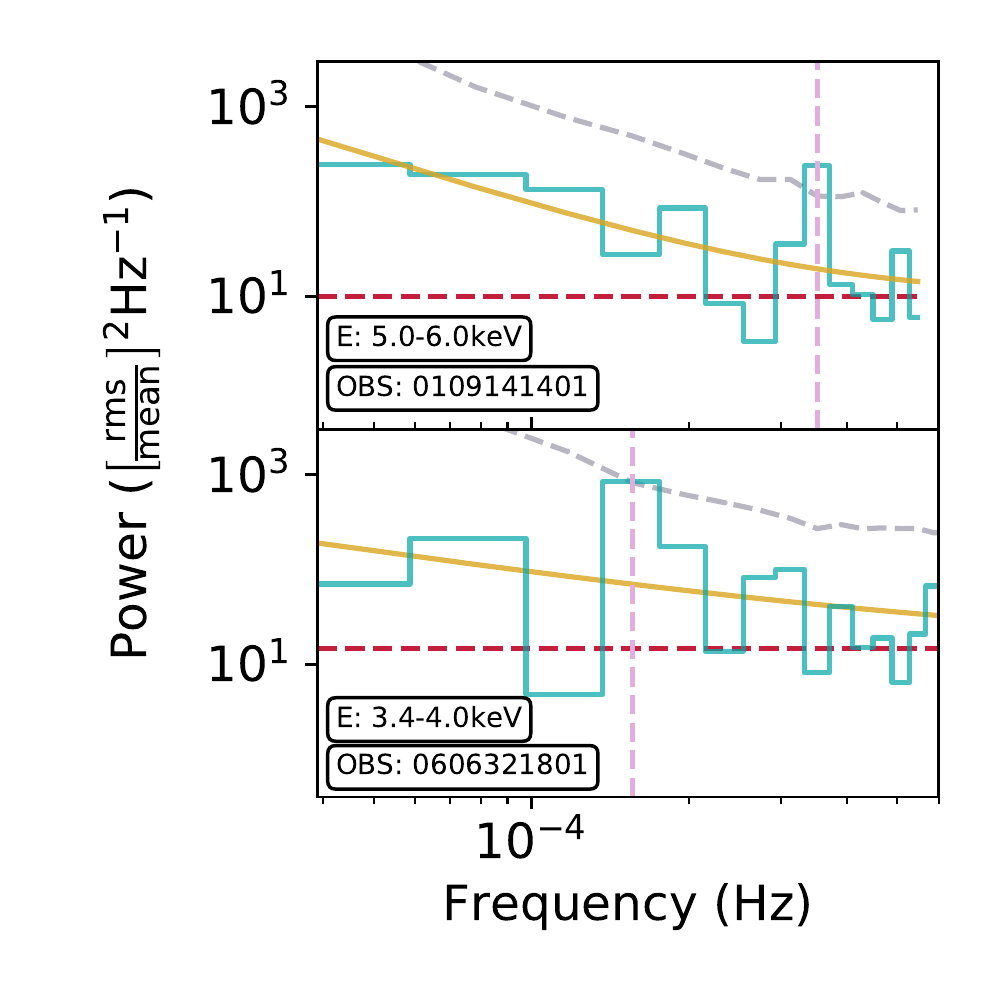}
    \caption{The highest significance ($\ge 3\sigma$) QPO candidates detected in NGC~4051 using the {\sc plc} model, with the corresponding frequency bin highlighted. We highlight those which are not validated by our further tests (\textbf{$\dagger$} -- see Section 3.2).}
    \label{fig:NGC_multiplots}
\end{figure}

Across 17 \textit{XMM-Newton} observations, we detect QPO candidates at $\ge 3\sigma$ significance in 3. However, we report only two candidates (see Table \ref{tab:QPOs}), at frequencies of $3.52\times10^{-4}$ Hz and $1.56\times10^{-4}$ Hz respectively. The former is shown as an example in Figure \ref{fig:Triple_plot_NGC}, and both candidates are compared directly in Figure \ref{fig:NGC_multiplots}. Notably, the QPO candidates in these observations are significant to $\ge 3\sigma$ in energy ranges of $4.6 - 7.6$\,keV and $3.2 - 5.4$\,keV respectively (as shown in Figure \ref{fig:Heatmap_panels}). Hence, we appear to observe both frequency and slight energy variation between these candidates. We regard the detection of a $\ge 3\sigma$ candidate in observation OBSID:0606320301, (with a peak significance of 0.9979, in the energy range 3.6 - 6.8\,keV, and at a frequency of $1.56\times10^{-3}$ Hz) to not be reliable as it sits in the highest Fourier frequency bin on the white noise cutoff.

\subsubsection{Further tests}

\medskip
\noindent{\bf Test(i)}:

\begin{itemize}
    \item OBSID:0109141401. A weak break at $\sim1\times10^{-4}$ Hz is identified in most energy ranges, and in each case the QPO candidate remains $\ge 3\sigma$. In a few energy ranges -- namely narrower bands such as 5.0 - 6.0~keV and 5.2 - 6.2~keV -- the observations now class as white noise dominated. 

\item OBSID:0606321801. A break is found consistently at $\sim1\times10^{-4}$ Hz, across all energy bands, close to the QPO candidate at $1.56\times10^{-4}$ Hz. The large offset of power above the break has led to identification of a QPO candidate when using the \textsc{plc} model, but as this is well described by a break, the candidate significance is now lowered to a borderline $3\sigma$ detection. Clearly this feature is data limited and requires further investigation.

\end{itemize}

\bigskip
\noindent{\bf Test(ii)}:

\begin{itemize}
    \item OBSID:0109141401. The power spectrum of this observation is well described with a lower index of -1.1. This leads to results almost identical to those from test (i), returning a QPO candidate well above the $3 \sigma$ threshold for those energy bins not deemed to be white noise dominated. 

\item OBSID:0606321801. Unlike OBSID:0109141401, the power spectrum is {\it not} well described with a lower index of -1.1, and the magnitude of the break is reduced near the QPO candidate frequency. In this case, the excess of power around the break is sufficient for the QPO candidate to remain detected at $> 3\sigma$ --  highlighting the importance of model selection.

\end{itemize}

\bigskip
\noindent{\bf Frequency-restricted search:} With two QPO candidate frequencies, it is difficult to motivate a single \textit{a-priori} frequency for a search across those observations without blind detections. Searching at the two frequencies reported in Table \ref{tab:QPOs}, we find no supporting evidence of a feature at $1.56\times10^{-4}$ Hz, but we do find a frequency-restricted $\ge 3\sigma$ candidate at $3.52\times10^{-4}$ Hz in OBSID:0606320301, strongest at 3.6 - 5.0~keV, similar to the case of OBSID:0109141401.

\bigskip
\noindent{\bf Previous claims:} NGC 4051 has been the subject of multiple X-ray variability studies over several decades, with \citet{1995PapadakisNGC4051qpo} reporting a broad QPO-like feature in \textit{EXOSAT} data at a frequency of $\sim 4\times10^{-4}$Hz, which appears stronger in their low-energy band $(0.05-2)$keV, than their medium-energy band $(2-10)$~keV. It should be noted that their methodology does not include the simulation of red noise  following \citet{1995TK}, but does consider fitting with complex models. They find a \textsc{plc + Gaussian} model provides a good description of the underlying power spectrum and QPO-like feature at low energies, with a highly significant improvement over a \textsc{plc} model. We note that the QPO candidate we observe in OBSID:0109141401 lies at a similar frequency ($3.52\times10^{-4}$ Hz), but is significant at harder energies within their medium-energy band (see Figure \ref{fig:Heatmap_panels}). 

\citet{1999GreenDoneNGC4051nonlinearvar} study two, $0.1-2$~keV \textit{ROSAT} observations of NGC 4051 and  suggest the possible presence of a QPO-like feature at frequencies $>\sim10^{-3}$Hz, but this is above the white noise limit for our observations. Lastly, \citet{2011VaughanNGC4051} conducted a variability analysis of NGC~4051 using a series of \textit{XMM-Newton} observations (with a total of $\sim570$~ks), across the full energy range of $0.2-10$~keV, and across a frequency range of $\sim 1\times10^{-4} - \sim 1\times10^{-2}$~Hz. They report no QPOs, although they do note that their analysis at harder energies ($2-10$~keV) provided tentative evidence for a possible feature at $\sim 4\times10^{-3}$~Hz.

\subsection{ARK 564}

ARK 564 is another well-studied NLS1, both on short (e.g. \citealt{Kara2017ARK564}) and long (e.g. \citealt{2006McHardyAGNScaling}) timescales, providing deeper insights into the complex nature of the broad-band noise, which we discuss in Section 5. 

\subsubsection{Initial {\sc plc} search}

\begin{figure}
    \centering
    \includegraphics[width=0.47\textwidth]{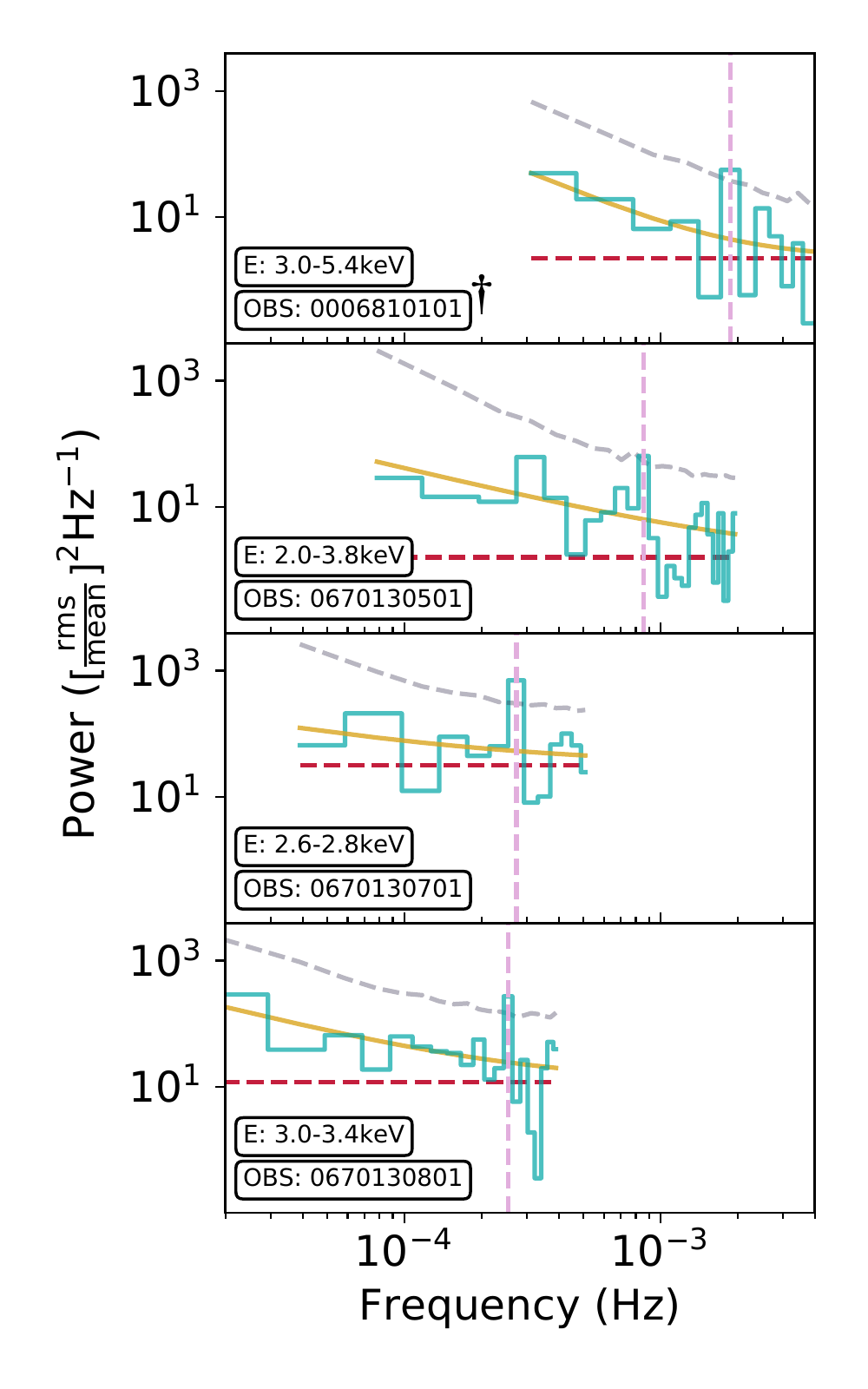}
    \caption{The highest significance ($\ge 3\sigma$) QPO candidates detected in ARK~564 using the {\sc plc} model, with the corresponding frequency bin highlighted. We note the apparent shift of the candidate to lower frequencies across each observation (and with time), where it appears to stabilise. We highlight those which are not validated by our further tests (\textbf{$\dagger$} -- see Section 3.2).}
    \label{fig:ARK_multiplot}
\end{figure}

We detect QPO candidates in 4 out of 11 \textit{XMM-Newton} observations. The nature of these candidates appears less stable  than in the other AGN in our sample, and Figure \ref{fig:ARK_multiplot} demonstrates the apparent movement of the QPO candidate towards lower frequencies with time. 

The first observation in which we find a $\ge 3\sigma$ outlier (OBSID:0006810101) is similar to the higher frequency QPO candidate we have located in IRAS~13224-3809, at a frequency of $1.88\times10^{-3}$ Hz. The longest segment in this observation is uncharacteristically short, at only 4.95~ks, but with a period of $\sim532$\,s we can still obtain a moderate quality factor of $Q \approx 6$. In OBSID:0670130501, we detect a QPO candidate at a frequency of $8.59\times10^{-4}$ Hz, prior to the observations of OBSID:0670130701 and OBSID:0670130801 in which we find QPO candidates at $2.73\times10^{-4}$ Hz and $2.54\times10^{-4}$ Hz respectively. The energy ranges in which these four QPO candidates are detected are largely consistent, being dominated by counts at $2-3\,$~keV. The full range of candidate significance as a function of energy is shown in Figure \ref{fig:Heatmap_panels}.

\subsubsection{Further tests}

\medskip
\noindent{\bf Test(i)}:

\begin{itemize}
    \item OBSID:0006810101. In requiring a break to occur at or before the QPO candidate, we find every energy bin in this observation to be considerably restricted in the number of available frequency bins below the white noise cutoff. As a result, we cannot discern the significance of this QPO candidate. As such, this detection is dependent on our selection of white noise cutoff, so as a result it difficult to discern from white noise and is brought into question.

\item OBSID:0670130501. We find a clear break at a frequency of $~3\times10^{-4}$ Hz above 2~kev, and find a break at $~5\times10^{-4}$ Hz when including energies below 2~keV. In the case of the former, we find the QPO candidate at $8.59\times10^{-4}$ Hz to increase in significance. In those energy bands where the break is located at $~5\times10^{-4}$ Hz, the significance of the QPO candidate is reduced, but remains around $3\sigma$.

\item OBSID:0670130701. We find a break at a frequency of $~8\times10^{-5}$ Hz in the 2.4 - 3~keV energy range, with the break moving to $~1\times10^{-4}$ Hz for 2.6 - 3.0~keV and $~2\times10^{-4}$ Hz for 2.6 - 3.2~keV -- with the QPO candidate at $2.73\times10^{-4}$ Hz. The QPO candidate remains $>3 \sigma$, but decreases in significance with increasing energy and a higher break frequency. We also find the peak significance band (2.6 - 2.8~keV) now classes as white noise dominated as the \textsc{bknplc} invokes a cutoff at a lower frequency.

\item OBSID:0670130801. This observation contains the strongest QPO candidate we have located in ARK~564. Once again, the break is highly energy dependent: a break is found at $\sim 1\times10^{-4}$ Hz in the 3.0 - 4.0~keV energy range, but a strong break is instead found at $\sim 6\times10^{-5}$ Hz for 5.0 - 10.0~keV, before the power spectrum flattens for energy bands $>$ 6~keV, resulting in a weak break. In each case, the QPO candidate is found above the $3\sigma$ threshold, although the significance lessens in energy bands $>$ 6~keV. 

\end{itemize}

\bigskip
\noindent{\bf Test(ii)}:

\begin{itemize}
    \item OBSID:0006810101. All energy bands are classified as white noise dominated assuming this model.

\item OBSID:0670130501. A lower index of -1.1 does not provide a 
good description for the break across the full range of energy bands. This model typically locates the break at $~6\times10^{-4}$ Hz and reduces the QPO candidate significance, although not below the $3\sigma$ threshold.

\item OBSID:0670130701. Again, a lower index of -1.1 does not allow for a good description of the periodogram, but the QPO candidate remains above the $3\sigma$ threshold.

\item OBSID:0670130801. As with the other observations, the periodograms are not well described with a lower index of -1.1. As a break is not preferred, most QPO candidates remain above the $3\sigma$ threshold, with the exception of energy bands $>$ 5.6~keV, in which the candidates lie at the white noise cutoff frequency.

\end{itemize}

\bigskip
\noindent{\bf Frequency-restricted search:} The observed variation in QPO candidate frequency means we have minimal \textit{a-priori} knowledge of a stable frequency for deeper searches, and therefore we exclude this step for ARK~564.

\bigskip
\noindent{\bf Previous claims:} \citet{2007IMcHardyArk564Lor} reported an excess of power in the 2.0 - 8.8\,keV band in OBSID:0206400101 at $\sim2\times10^{-4}$\,Hz -- similar to two of our QPO candidates in OBSID:0670130701 and OBSID:0670130801 -- but only at a 90\% significance. Our analysis of this observation does not reproduce the feature, as the longest continuous segment of this $\sim101$\,ks observation is only $\sim21$\,ks, severely limiting our frequency range.

\subsection{MRK 766}

MRK~766 is again well-studied (e.g. see \citealt{2007MarkowitzMRK766}), with nine \textit{XMM-Newton} observations in the archive at the time of writing.

\subsubsection{Initial {\sc plc} search}

\begin{figure}
    \centering
    \includegraphics[width=0.47\textwidth]{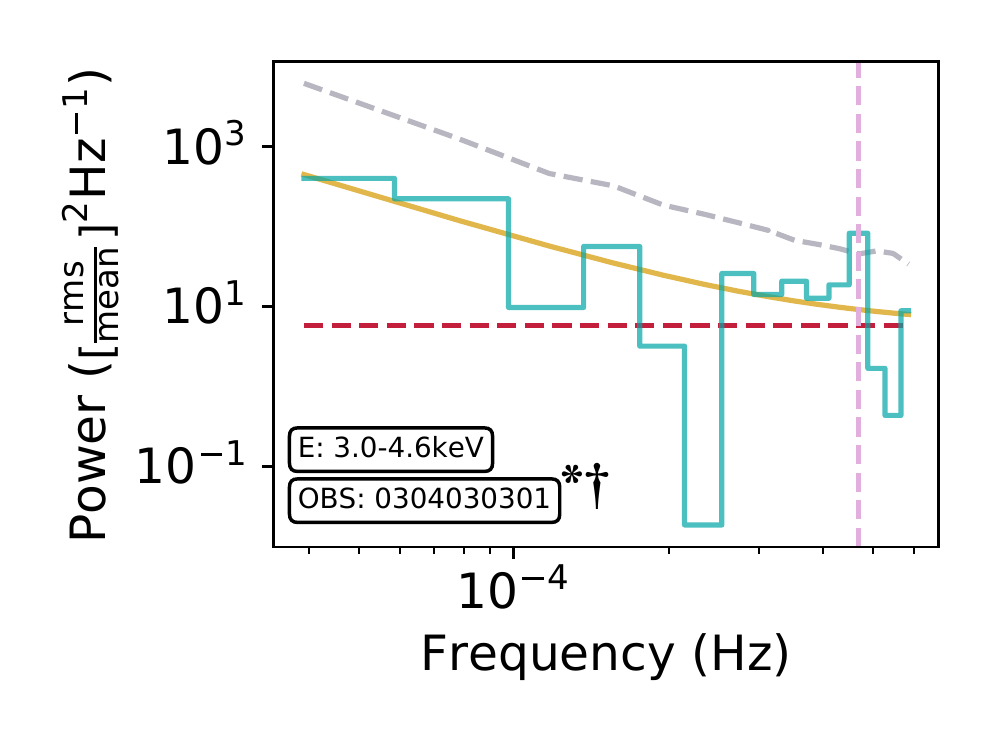}
    \caption{The highest significance -- borderline -- QPO candidate detected in the single observation of MRK~766 using the {\sc plc} model, with corresponding frequency bin highlighted.}
    \label{fig:MRK_singleplot}
\end{figure}

We report a single QPO candidate at $\ge 3\sigma$, in a single energy bin, $3.0 - 4.6$\,keV, as shown in Figure \ref{fig:MRK_singleplot}. Figure \ref{fig:Heatmap_panels} demonstrates how the QPO candidate significance drops below the $3\sigma$ threshold in surrounding energy bins. At a frequency of $4.69\times10^{-4}$ Hz, the QPO candidate frequency is similar to that of the other candidates we have located in our sample.

As our significance tests are based on an underlying stochastic background, it is quite possible we may determine an outlier to lie very close to $3\sigma$ (peaking above or dipping below this threshold under repeated tests). For our analysis of MRK 766, this occurs for our only candidate. This borderline $3\sigma$ detection, coupled with a lack of supporting evidence from other observations (e.g. in IRAS 13224-3809 and 1H~0707-495 we have found multiple QPO candidates across multiple observations at the same frequency across similar energy ranges) means we remain sceptical of this detection, however we report the findings to allow a comparison to be made to future studies.

\subsubsection{Further tests}

\medskip
\noindent {\bf Test(i)}:

\begin{itemize}
    \item OBSID:0304030301. We find a tentative break at $\sim1\times10^{-4}$ Hz and white noise cutoff at $\sim5\times10^{-4}$ Hz, which heavily restricts the number of frequency bins such that the periodograms are defined as white noise dominated.

\end{itemize}

\bigskip
\noindent{\bf Test(ii)}:

\begin{itemize}
    \item OBSID:0304030301. With a frozen lower index of -1.1 we obtain identical results to test (i).

\end{itemize}

\bigskip
\noindent {\bf Frequency-restricted search:} 

For completeness, we assume an \textit{a-priori} QPO candidate frequency of $4.69\times10^{-4}$ Hz, and searched within the other observations, finding a restricted-frequency $>3 \sigma$ candidate in OBSID:0096020101, strongest in the 0.7 - 1.2~keV band -- notably at a lower energy range than in OBSID:0304030301.

\bigskip
\noindent{\bf Previous claims:} We note the previous claims of a strongly significant QPO in a single observation of MRK 766 (OBSID:0304030601) identified through a non-Fourier wavelet method (\citealt{Zhang2017MRK766}). However, we do not find a significant feature at their reported frequency of $1.55\times10^{-4}$ Hz.

\subsection{Observations as free trials}

The significance values we have quoted for each QPO candidate rely on the assumption that each observation is independent and the number of free trials is limited to only the frequency range of interest, per observation. Our motivation for not treating each observation as being a free trial stems from the observation that certain spectral states, connected to the accretion rate, seem to be necessary for such QPOs to appear (e.g. \citealt{2014Alston5}).

We can instead employ the most conservative perspective and assume each observation (200 in total) to be a free trial -- regardless of data quality -- and obtain the most stringent QPO candidate significance.

In the case of QPO candidates present at a single frequency in a single observation, such as PG 1244+026 and NGC 4051, we find the significance values fall to $<3 \sigma$. In the case of NGC 4051 OBSID:0109141401, this is reduced to the $\sim 2\sigma$ level, while PG 1244+026 OBSID:0675320101 falls to $<2 \sigma$. However we note that such decreases are overestimates of the impact (as we have assumed the data quality to be equivalent in all observations).

For those AGN with QPO candidate detections at the \textit{same frequency} across multiple observations, we consider both the more stringent `look elsewhere' effect set against the conditional probability of detecting a QPO candidate at the same frequency. Combining the two leads to the QPO candidate in 1H~0707-495 being at $> 6 \sigma$ significance, while the IRAS 13223-3809 QPO candidate at a frequency of $9.77 \times 10^{-5}$Hz may similarly be found at $> 6 \sigma$ -- although this drops to $> 4 \sigma$ if excluding the detection in OBSID:0792180301, where the broad-band noise is not well constrained. In the cases of ARK~564 and RE~J1034+396, our global significance values for some candidates of interest reach $GS = 1.0000$, whereby we reach a precision limit for $10,000$ MC simulations. If instead we consider these individual global significance values to be similar to those  observed in 1H~0707-495, we find the $2.54 - 2.73 \times 10^{-4}$Hz candidate in ARK~564 to be $> 4\sigma$, whilst RE~J1034+396 reaches the $>7 \sigma$ level, with four observations all at equivalent frequencies. We note that these values in themselves are conservative, as they consider only $> 3\sigma$ global significance values in Table \ref{tab:QPOs}, and no supplementary evidence from observations with lower significance candidates.

\section{Additional uncertainty in the broad-band noise}

We have chosen to focus on relatively simple models for the broad band noise as these are typically a good description in the frequency range we are studying (e.g. \citealt{2012GonzMartinVaughan}). However, there is the potential for further underlying complexity in the broad band noise and leakage which can effect our QPO candidate significance claims and which we explore here. 

\subsection{The impact of the white noise cutoff}

Determining the frequency at which the underlying white noise (Poisson noise) within the observation becomes dominant over the intrinsic red noise is subject to inherent uncertainty in any fitting of the broad-band noise. In our approach to defining the null hypothesis (Section 3.1.1) -- the QPO candidate is removed from the periodogram and the broad-band noise is re-fitted (see \citealt{2005Vaughan}), but the white noise cutoff frequency is left {\it unchanged} from our fit where the QPO candidate is included. Defining this frequency is a somewhat complex issue as the periodogram is by definition only one realisation of the underlying power spectrum (and so would be best represented by a probability distribution) and the position of the cutoff naturally has an uncertainty, correlated with those uncertainties on the normalisation and power-law index (such that a different draw from the covariant distribution -- Figure \ref{fig:GMM_plot} -- produces a different cutoff). As we are required to select a single cutoff frequency (otherwise we introduce an uneven bias into our significance testing), we opted to retain the value from our initial fit but now explore the impact of this choice.

Re-fitting the broad-band noise with the QPO candidate removed tends to reduce the white noise cutoff frequency. The result is that bins at high frequency may then be outside of our range of interest. As we also require 13 Fourier bins (which includes the QPO candidate bin -- although this frequency is ignored when the null hypothesis is defined) for fitting purposes, this can have the knock-on effect of removing energy bands from our study. The combination of these two effects can remove QPO candidates either because they now fall into the white noise dominated region or because there are too few bins for reliable fitting. We investigated these effects by setting the white noise cutoff frequency from the {\sc plc} model applied to the data without the QPO candidate. We find that $\sim39\%$ of the $> 3\sigma$ energy bands are now white noise dominated (with $<$ 13 frequencies in the periodogram, or 12 bins when excluding the QPO candidate bin), a change driven by the removal of those narrow energy bands with relatively low count rates and so large amounts of white noise, which were close to this boundary in our original analysis. Despite the loss of numerous energy bands, the majority of our reported $> 3\sigma$ QPO candidates remain, with a few exceptions, namely: IRAS 13224-3809 OBSIDs: 0780561301 $(\nu = 7.81\times10^{-5}$Hz only), 0792180301, 0792180601; 1H 0707-495 OBSID:0653510301, ARK 654 OBSID:0006810101, MRK 766 OBSID:0304030301, and perhaps most notably: RE~J1034+396 OBSID:0675440101 (as well as almost all energy bins of OBSID:0655310201). It should be noted that a number of these observations were already borderline/close to the white noise threshold.

\begin{figure}
    \centering
    \includegraphics[width=0.47\textwidth]{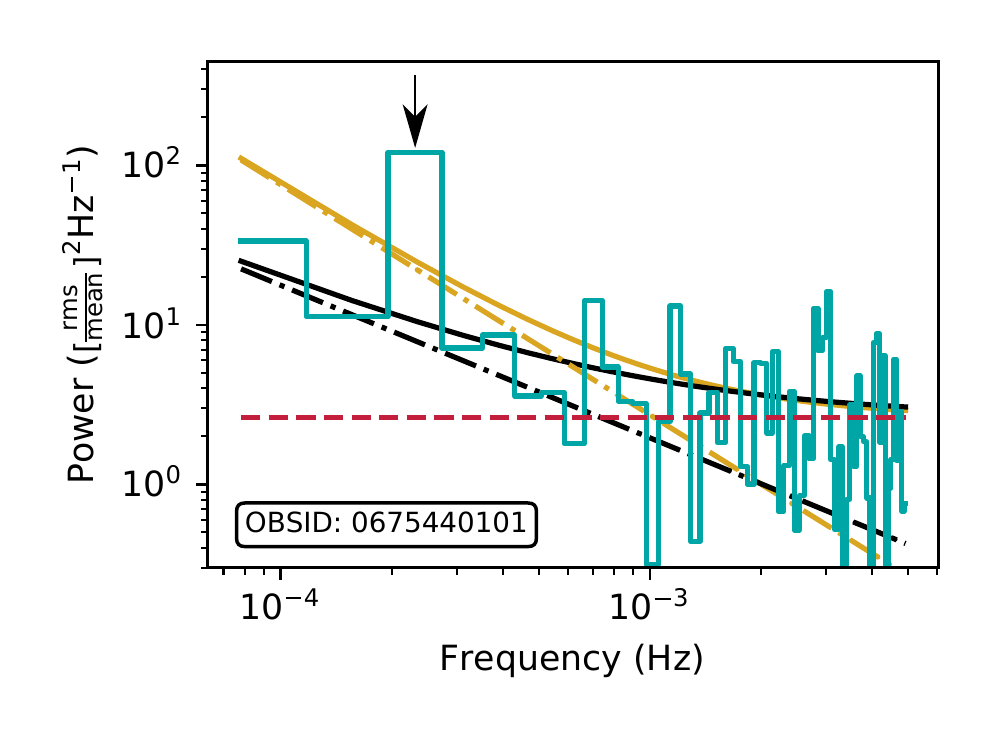}
    \caption{Example of the impact of the white noise cutoff selection via our two approaches (see Section 5.1), for the established QPO in RE~J1034+396, OBSID: 0675440101 (within the most significant energy band, see Table \ref{tab:QPOs}). We show two \textsc{plc} model fits, one (gold) including the QPO candidate bin, and the other (black) without the QPO candidate bin. The respective white noise cutoff frequencies in each case are determined to be the point where the underlying  \textsc{plc} model (dot-dashed) intersects the constant white noise level. As shown here, excluding the QPO candidate in our fitting results in the frequency of the white noise cutoff being lowered on average by $\sim2 - 3$ bins. As we demonstrate in this example, we then risk reducing the number of bins in our periodogram to below the minimum requirement we have set for modelling the broad band noise -- this would lead us to exclude this established and accepted QPO.}
    \label{fig:WNthreshold}
\end{figure}

We note that, although setting the white noise cutoff to the higher of the two frequencies is a more conservative approach, it also has the potential to remove real signals, as is evident for the case of RE~J1034+396 (see Figure \ref{fig:WNthreshold}). In a similar fashion, the QPO candidate of 1H~0707-495 in OBSID:0653510301, is discarded by this more stringent analysis but is still observed in OBSID:0653510501 {\it at the same frequency}, and so we find we can lose high probability detections. 

\subsection{The effect of windowing/red noise leak}

\begin{figure}
    \centering
    \includegraphics[width=0.47\textwidth]{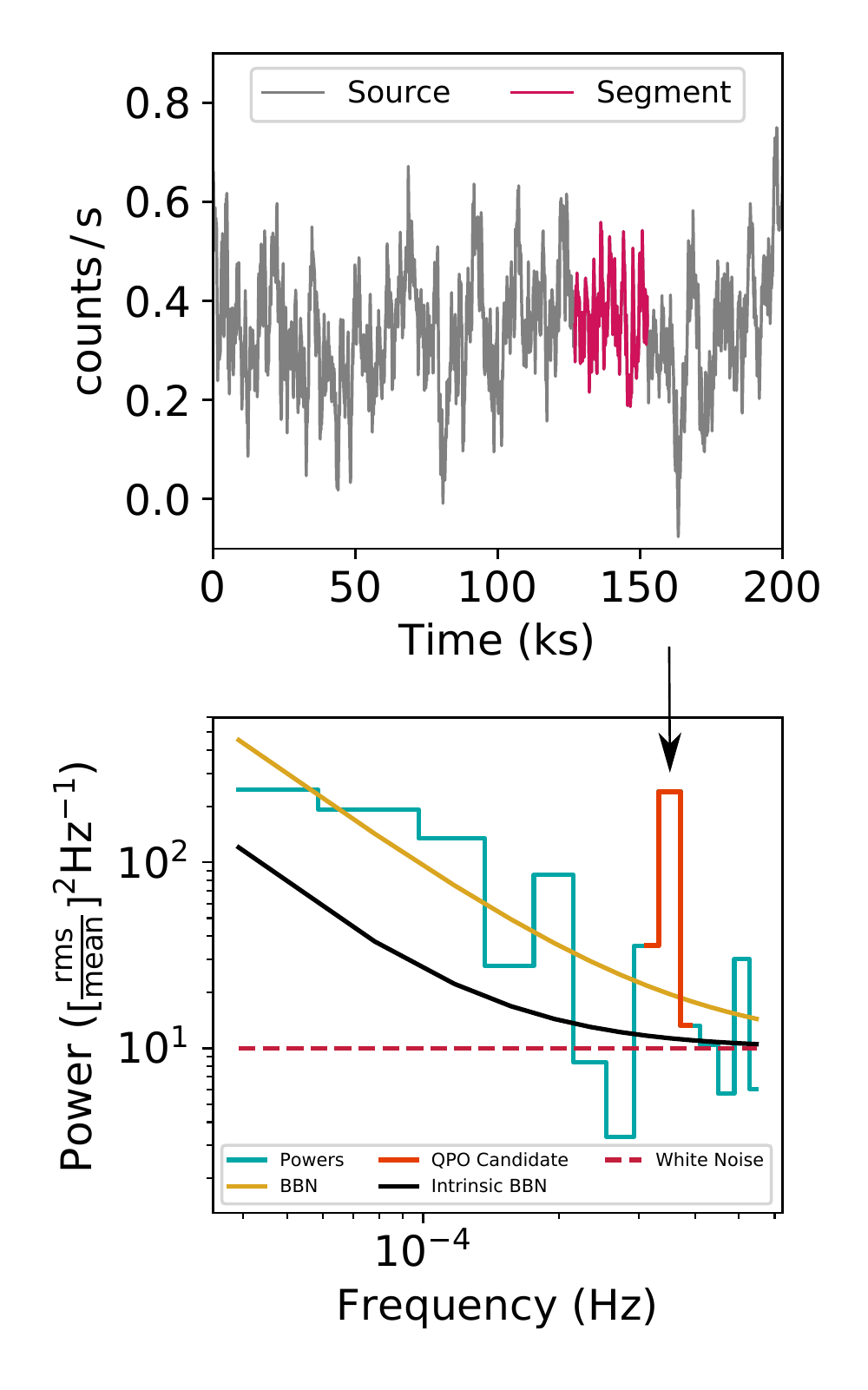}
    \caption{Top: example of a simulated light curve for NGC 4051 (OBS ID:0109141401), with a highlighted, randomly selected segment of length equal to the longest observed continuous segment of our real data. Bottom: observed periodogram (QPO highlighted), with best fitting {\sc plc} model and the intrinsic, pre-windowed {\sc plc} model.}
    \label{fig:Window_plot}
\end{figure}

As we are only selecting a relatively small frequency range compared to the full extent of the underlying power spectrum (see e.g. \citealt{2002UttleyBBN}, \citealt{2006McHardyAGNScaling}), we may expect some leakage of power into our observed band \citep[see][]{2012GonzMartinVaughan}. In Fourier space this is equivalent to convolving the Fourier transform of the window function (the Fej{\'e}r kernel) with the underlying power spectrum \citep{1988VandK}. As a result, the intrinsic broad-band noise may not be well represented by that which we measure in our periodogram. We explore the impact of leakage on those periodograms in which we have significant QPO candidate detections, by simulating a lightcurve based on an underlying power-law with an index of -2 following \citet{1995TK} but extending back to 10$^{-5}$~Hz, i.e. to lower frequencies than we typically observe down to, and then apply a window equivalent to our segment selection. To obtain an appropriate normalisation for our input power-law, we first create very long simulated lightcurves (far longer than the real data) from a power-law with an index of -2 and trial normalisation stepping over log space. From these lightcurves we select a random segment of the same length as our real data and minimise the log-likelihood when compared to our {\it observed} periodogram (by minimising the median of the distribution of $S$ values). Using this best-fitting normalisation -- which is different for each QPO candidate-detected observation -- we simulate 10,000 fake lightcurves, apply our window, obtain the periodogram, and search for false outliers. We note this to be a somewhat extreme test as it is quite likely that a break to to flatter indices occurs before such low frequencies are reached (see e.g. \citealt{2007IMcHardyArk564Lor}) or that the power-spectrum itself is flatter ($\beta >$ -2) as we move to higher energies (\citealt{2020JinREJ}; Ashton et al. in prep.) where our QPO candidates are mostly located. In Figure \ref{fig:Window_plot} we present an example of our test for the impact of windowing as applied to NGC~4051 (OBSID:0109141401), where we plot the intrinsic noise needed to create the observed noise.

We apply the above test to only the most significant energy-resolved QPO candidates (as listed in Table \ref{tab:QPOs}) and find that 8 out of the 24 candidates remain above the $3\sigma$ threshold. These include RE~J1034+396 (OBSIDs: 0506440101 and 0675440201), IRAS~13224-3809 (OBSIDs: 0780561401 and 0792180501), NGC~4051 (OBSID: 0109141401) and ARK564 (OBSIDs: 0006810101, 0670130701 and 0670130801). The remainder lie above a $\sim 98\%$ threshold, i.e. an upper $2\sigma$ level. This is a clear indication that under extreme conditions it is possible for red noise leak to have an impact on QPO detection. However, we note that this may not rule out the now $< 3\sigma$ candidates as, in many cases (e.g. RE~J1034+396), we find the QPO candidate frequencies to remain similar across multiple observations, which is hard to reproduce through fake signals generated by an underlying stochastic process. 

\subsection{Complex broad-band noise models}

\begin{figure*}
    \centering
    \includegraphics[width=\textwidth]{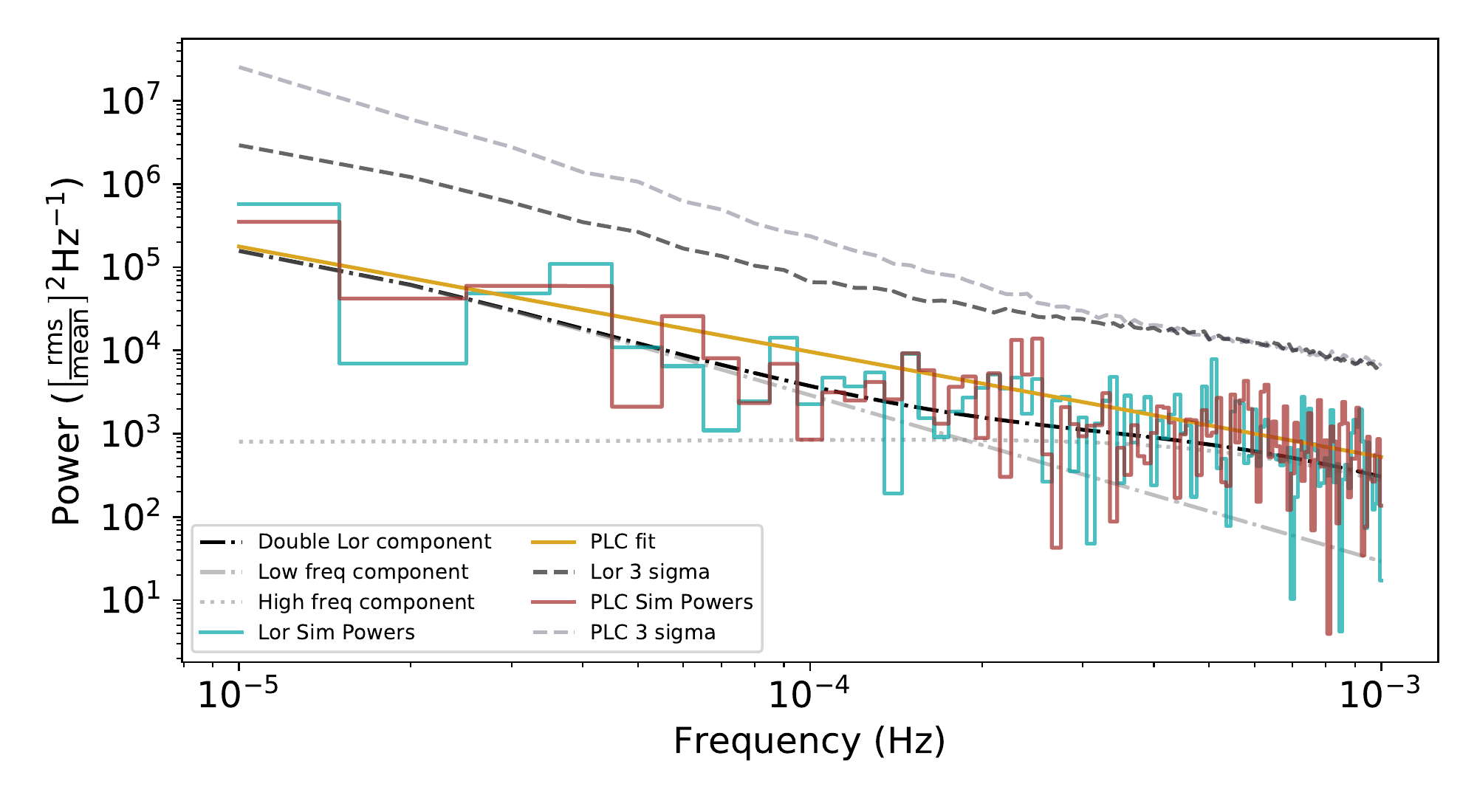}
    \caption{Simulated power spectra from 10$^{-5}$ - 10$^{-3}$~Hz, generated from the high energy, double Lorentzian model for ARK~564 \citep{2007IMcHardyArk564Lor} and a \textsc{plc} model fitted to the data generated from this. The model choice has a notable impact on the $3\sigma$ significance contours (\citealt{2005Vaughan}). In this case, should we assume a \textsc{plc} model but a double Lorentzian model is the correct description of the broad-band noise, we would potentially underestimate QPO candidate significance at low frequencies. }
    \label{fig:Lorentzianplot}
\end{figure*}

A natural extension to a test using a {\sc bknplc} model is one in which the underlying noise is described by a series of Lorentzians (e.g. \citealt{2007IMcHardyArk564Lor}). This makes sense physically where each radius acts as a low pass filter with variations suppressed on a local viscous timescale (e.g. \citealt{2001Churazov}; \citealt{2016Ingram}), although in reality these are well known to be coupled \citep{2017Uttleycoupling}. Whilst we do not have {\it a-priori} knowledge of the nature of the Lorentzians that might describe the shape of the broad band noise in any of our energy-dependent observations, we can test the impact such a model might have on QPO significance tests more generally. Assuming the high energy Lorentzians which describe the data of ARK~564 from \citet{2007IMcHardyArk564Lor} without white noise, we create fake periodograms down to 10$^{-5}$~Hz, a distribution of $R$ values, and obtain the free-trial-corrected 3$\sigma$ contours (see \citealt{2005Vaughan}). We then proceed to fit the average of these simulated periodograms with our {\sc plc} model (minus white noise) and, using the best-fitting value for the index and normalisation, simulate again to obtain another set of 3$\sigma$ contours. As shown in Figure \ref{fig:Lorentzianplot}, by comparing the two sets of contours, we can see that, over much of the frequency range, the significance is under-estimated when a power-law model is used as an approximation, although the significance contours are comparable as the higher frequency Lorentzian starts to dominate the broad band noise. 

Should a multiple Lorentzian model be a more appropriate description of the underlying noise for all of our observations at high energies, then the impact on our QPO candidate significance is expected to be relatively minor. However, given the lack of constraints on the nature and energy-dependence of any Lorentzians underlying our data, we stress that this is an open question and reiterate that accurately modelling the shape of the noise is vital for any robust significance claims (see \citealt{2016FalsePeriodicities} for more discussion on this topic).

\section{Discussion}
To-date, the presence of AGN X-ray QPOs across multiple observations has been limited to strong detections in the NLS1 AGN, RE~J1034+396 \citep{2008Nature}, with a single claim in MS 2254.9-3712 \citep{2015AlstonMS2254}, a number of tentative claims in other NLS1s, and signals in TDEs (see \citealt{2019QPOsinTDEs}).  Here we have performed the first energy-resolved search for QPO candidates within 38 AGN, comprising 200 \textit{XMM-Newton} observations. We initially find 24 QPO candidates across 7 AGN, and subject to further tests, report the detection of 16 QPO candidates across 6 AGN (we exclude MRK~766 from this category -- see Section 4.7).

Our results include the QPO in RE~J1034+396 which we used to corroborate our approach, as well as confirming signals in IRAS~13224-3809, 1H~0707-495 and ARK~564 albeit it in some cases at different frequencies to those quoted in the past. In addition, we provide details of the energy dependence and the repeat occurrence of these signals across observations. This last point is important - in many cases (notably IRAS 133224-3809 and 1H~0707-495) the QPO candidates appear at the same or very similar frequencies (see Table \ref{tab:QPOs}); it is very difficult to create such repeat signals by chance alone, increasing the likelihood that they are genuine (see Section 4.8, where we reach significance levels in excess of $>6 \sigma$ for 3 AGN). We also note that, in most cases, the quality factor and rms of the signal is large (Q $>$ 10 and FRMS $\gtrsim$ 10\%) in keeping with those established QPOs in RE~J1034+396.

A possible limitation in our approach to estimating the significance of our QPO candidates, is our data-led maximum likelihood search for a break in the power spectrum, which does not incorporate any prior expectations we have for the position of the break (e.g. \citealt{2007IMcHardyArk564Lor}; \citealt{2012GonzMartinVaughan}). At present, we merely {\it force} a break into the periodogram as a rigorous test of the QPO candidate significance (see Section 3.2); the occasional change in QPO candidate significance then underlines the need for caution when claiming the presence of a QPO, even when the broad-band noise model is statistically preferred. In future, and with data which extends to lower frequencies, we will be able to perform more stringent tests. We note that, whilst our approach may appear to discard information prior to our fitting, the present lack of a well-constrained probability distribution describing the position and energy dependence of the break prevents such an approach. Given the situation of poorly motivated priors and somewhat limited data, we have chosen to follow a frequentest approach, although Bayesian inference may allow for a more accurate assessment of the presence of breaks in future. Bearing in mind the above caveats and our warnings about the uncertainty in the broad-band noise, we proceed to discuss the QPO candidates our analysis has identified, and the possible unifying features in our sample. 

RE~J1034+396 is well known for having a strong soft excess \citep{2002PuchnarewiczREJ, 2006CasebeerREJ, 2007MiddletonPGQSRs, 2009MiddletonSoftExcess}, likely a result of extremely high accretion rates, which has led to comparisons being made between its QPO and the 67~Hz HFQPO observed in GRS~1915+105 when accreting at high Eddington fractions (\citealt{1997MorganGRS1915}; \citealt{2009UedaGRS1915}; \citealt{2010MiddletonBHBAnalogy}, \citealt{2020JinREJ}). More generally, HFQPOs are only seen in BHBs in their very high state (see \citealt{Remillard2006XRBproperties}), when the mass accretion rates can reach significant fractions of their Eddington limit (e.g. see \citealt{1995Nowak}). It is therefore reasonable to postulate that the HFQPO  mechanism is intrinsically tied to these high mass accretion rates \citep{2006BlaesArrasFragile}. 
 
Of our sample of AGN containing QPO candidates, RE~J1034+396, IRAS 13224-3809, PG~1244+026, 1H~0707-495 and ARK~564 have all been reported accreting at close to their Eddington limits \citep{2016DoneJin, 2020AlstonIRASreverbmap, Kara2017ARK564, 2012DoneIntrinsicDiscEmission} as is also the case for MS~2254.9-3712 \citep{2015AlstonMS2254}.  

Indeed, NGC~4051 may be the only exception, accreting at $\sim0.1$ Eddington \citep{2000PetersonNGC4051, 2013AlstonNGC4051}. Whilst the uncertainty involved in mass accretion rate determination means we cannot be completely certain of the true nature of these sources, there could well be a link between the high accretion rates in these AGN (as well as in TDEs) and the QPO candidates we have found.

Naturally, high accretion rates (and therefore high luminosities) could result in some level of selection bias -- our ability to detect a QPO relies on high signal-to-noise ratio data and so we may be more likely to detect signals in predominantly bright, high accretion rate AGN. We have been unable to test this in our present sample due to the uncertainty in SMBH mass measurements and therefore accretion rate estimates. It is quite possible however, that QPOs in AGN are far more commonplace than previously thought, but the present difficulty in obtaining the required data quality to probe a sufficient frequency range, biases our study towards intrinsically bright, or sufficiently local AGN. In future, new, high-throughput instruments such as {\it STROBE-X} and {\it Athena} will be able to probe many more AGN to higher frequency ranges.

The energy bands in which we have found significant QPO candidates vary between sources, but remain largely consistent between observations of the same source and, with the exception of RE~J1034+396, the detections are most significant in bands above 2\,keV. This was also found by \citet{2015AlstonMS2254} in their investigation of MS~2254.9-3712, where the QPO is detected at $2-5\,$keV and provides further compelling evidence that the hard energy components (i.e. the coronal emission and reflection), of the AGN energy spectrum appear to be central to QPO production. The energy-dependence of detection significance also indicates why previous studies (e.g. \citealt{2012GonzMartinVaughan}) have been unlikely to detect these QPO candidates, as they do not appear in the typically studied full bandpass $0.3-10\,$keV, nor generic soft, $0.3-2\,$keV, or hard, $2-10\,$keV bands. In a forthcoming paper we will investigate the full energy-dependence and detailed nature of these candidates to search for  confirmatory evidence of their QPO (i.e. non-broad-band noise) nature, and thereby hope to better understand the QPO production mechanism in AGN.

The QPO frequency is naturally an important parameter in understanding the accretion flow (e.g. as a result of motion of the flow or characteristic timescale of an instability). Our prior knowledge of QPO frequencies in this regard was somewhat limited to the $\sim2.5\times10^{-4}$\,Hz detection in RE~J1034+396 \citep{2008Nature} and the $\sim1.5\times10^{-4}$\,Hz detection in MS~2254.9-3712 \citep{2015AlstonMS2254}. Here we have found QPO candidates with similar frequencies in 1H~0707-495 and PG~1244+026, whilst NGC~4051 and IRAS~13224-3809 have QPO candidates with comparable frequencies, but with greater variation between observations. 

IRAS~13224-3809 is of particular interest for both the number of $\ge 3\sigma$ candidates and the consistency of their frequencies, with many close to $\sim1\times10^{-4}$\,Hz. The presence of {\it possible} harmonic features (the nature of which we will explore in more detail in the near future) is also unique amongst all of our observations, although this may simply be a consequence of being able to probe to higher frequencies due to the higher count rates in this AGN. As further indicated by the higher frequency QPO candidate detected in ARK 564 ($\sim1\times10^{-3}$\,Hz), being able to probe to high, non-white noise-dominated frequencies (especially at higher energies) may reveal many more QPO candidates. 

ARK~564 is rather unusual amongst our sample, with a QPO candidate  frequency decreasing with time, before an apparent stabilisation around $\sim2\times10^{-4}$\,Hz. This may be due to the changing frequency range we can probe, as the lowest frequency and white noise cutoff frequency also evolve across observations. Conversely, it may be the result of the physical process creating a genuine QPO changing with time, potentially settling, as suggested in the case of RE~J1034+396 \citep{2008Nature}. As noted earlier, these results assume a simplified description of the underlying broad-band noise and, for ARK 564 specifically, previous modelling has identified the presence of breaks and Lorentzians (e.g.\citealt{2002PapadakisArk564Break}, \citealt{2007IMcHardyArk564Lor}), which in one observation occurs close to our QPO candidate frequency (OBSID:0006810101). As we point out via simulations, it is unlikely that modelling the broad-band noise with Lorentzians could lead to an overestimate of QPO significance, however, as we have demonstrated, care should be taken when assuming the shape of the broad-band noise to be simple.

Although we will probe the detailed nature of these QPO candidates in a follow-up paper, we note that, should they be analogous to those HFQPOs in BHBs (e.g. \citealt{2010MiddletonBHBAnalogy}) then a simple inverse-mass relation might be expected for the QPO frequency. Unfortunately, the SMBHs in our sample are of comparable mass (see Table \ref{tab:QPOs}), and the present uncertainty in the mass measuring techniques renders any correlation to be dominated by uncertainty.

\section{Conclusion}

The search for X-ray QPOs in AGN has been an ongoing challenge within the field for a number of decades \citep{Vaughan05WhereAretheQPOs?} with multiple studies claiming QPO detections (e.g. \citealt{Vaughan05WhereAretheQPOs?}). To-date, only two AGN (non-TDEs) have signals which have been widely accepted as genuine \citep{2008Nature, 2015AlstonMS2254}. Following \citet{2015AlstonMS2254} we have analysed energy-dependent periodograms following a robust statistical method and have located significant ($\ge 3\sigma$) QPO candidates in five additional AGN. 

A number of these candidates appear at the same or similar frequencies across multiple observations, further diminishing the likelihood that they are spurious in nature. However, whilst our approach is robust, there naturally remains uncertainty over the exact shape of the broad-band noise -- although we don't anticipate this affecting our results, we will investigate the nature of our candidates in the near future to establish if they demonstrate QPO-like properties.

Our robust detection of QPO candidates using a rigorous approach not only expands the known sample size of such signals, but opens up the possibility of exploring the QPO formation mechanism in new, remarkable detail (Ashton et al. in prep) and implies that new missions (e.g. {\it Athena} and {\it STROBE-X}) will more regularly detect such features.

\section*{Acknowledgements}

DA and MM thank the anonymous referee for their useful suggestions. This research was possible through an STFC studentship. MM appreciates support via an STFC Ernest Rutherford Fellowship. We thank Will Alston, Simon Vaughan, Phil Uttley, James Leftley, John Paice, Peter Boorman and Adam Hill for helpful discussion. We also reference the use of python libraries which were essential in this analysis: Numpy \citep{Numpy}, Scipy \citep{Scipy}, and Matplotlib \citep{Matplotlib}.

\section*{Data Availability}

The data underlying this article are freely accessible in the public HEASARC archives (https://heasarc.gsfc.nasa.gov/).




\bibliographystyle{mnras.bst}
\bibliography{bibliography.bib}  

\begin{thebibliography}{}
\makeatletter
\relax
\def\mn@urlcharsother{\let\do\@makeother \do\$\do\&\do\#\do\^\do\_\do\%\do\~}
\def\mn@doi{\begingroup\mn@urlcharsother \@ifnextchar [ {\mn@doi@}
  {\mn@doi@[]}}
\def\mn@doi@[#1]#2{\def\@tempa{#1}\ifx\@tempa\@empty \href
  {http://dx.doi.org/#2} {doi:#2}\else \href {http://dx.doi.org/#2} {#1}\fi
  \endgroup}
\def\mn@eprint#1#2{\mn@eprint@#1:#2::\@nil}
\def\mn@eprint@arXiv#1{\href {http://arxiv.org/abs/#1} {{\tt arXiv:#1}}}
\def\mn@eprint@dblp#1{\href {http://dblp.uni-trier.de/rec/bibtex/#1.xml}
  {dblp:#1}}
\def\mn@eprint@#1:#2:#3:#4\@nil{\def\@tempa {#1}\def\@tempb {#2}\def\@tempc
  {#3}\ifx \@tempc \@empty \let \@tempc \@tempb \let \@tempb \@tempa \fi \ifx
  \@tempb \@empty \def\@tempb {arXiv}\fi \@ifundefined
  {mn@eprint@\@tempb}{\@tempb:\@tempc}{\expandafter \expandafter \csname
  mn@eprint@\@tempb\endcsname \expandafter{\@tempc}}}

\bibitem[\protect\citeauthoryear{{Alston}, {Vaughan}  \& {Uttley}}{{Alston}
  et~al.}{2013}]{2013AlstonNGC4051}
{Alston} W.~N.,  {Vaughan} S.,   {Uttley} P.,  2013, \mn@doi [\mnras]
  {10.1093/mnras/stt1391}, \href
  {https://ui.adsabs.harvard.edu/abs/2013MNRAS.435.1511A} {435, 1511}

\bibitem[\protect\citeauthoryear{{Alston}, {Markevi{\v c}i{\= u}t{\.e}},
  {Kara}, {Fabian}  \& {Middleton}}{{Alston} et~al.}{2014}]{2014Alston5}
{Alston} W.~N.,  {Markevi{\v c}i{\= u}t{\.e}} J.,  {Kara} E.,  {Fabian} A.~C.,
   {Middleton} M.,  2014, \mn@doi [\mnras] {10.1093/mnrasl/slu127}, \href
  {http://ukads.nottingham.ac.uk/abs/2014MNRAS.445L..16A} {445, L16}

\bibitem[\protect\citeauthoryear{{Alston}, {Parker}, {Markevi{\v c}i{\=
  u}t{\.e}}, {Fabian}, {Middleton}, {Lohfink}, {Kara}  \& {Pinto}}{{Alston}
  et~al.}{2015}]{2015AlstonMS2254}
{Alston} W.~N.,  {Parker} M.~L.,  {Markevi{\v c}i{\= u}t{\.e}} J.,  {Fabian}
  A.~C.,  {Middleton} M.,  {Lohfink} A.,  {Kara} E.,   {Pinto} C.,  2015,
  \mn@doi [\mnras] {10.1093/mnras/stv351}, \href
  {http://ukads.nottingham.ac.uk/abs/2015MNRAS.449..467A} {449, 467}

\bibitem[\protect\citeauthoryear{{Alston} et~al.,}{{Alston}
  et~al.}{2019}]{2019AlstonIRASvariability}
{Alston} W.~N.,  et~al., 2019, \mn@doi [\mnras] {10.1093/mnras/sty2527}, \href
  {https://ui.adsabs.harvard.edu/abs/2019MNRAS.482.2088A} {482, 2088}

\bibitem[\protect\citeauthoryear{{Alston} et~al.,}{{Alston}
  et~al.}{2020a}]{2020Alston}
{Alston} W.~N.,  et~al., 2020a, \mn@doi [Nature Astronomy]
  {10.1038/s41550-019-1002-x}, \href
  {https://ui.adsabs.harvard.edu/abs/2020NatAs.tmp....2A} {p.~2}

\bibitem[\protect\citeauthoryear{{Alston} et~al.,}{{Alston}
  et~al.}{2020b}]{2020AlstonIRASreverbmap}
{Alston} W.~N.,  et~al., 2020b, \mn@doi [Nature Astronomy]
  {10.1038/s41550-019-1002-x}, \href
  {https://ui.adsabs.harvard.edu/abs/2020NatAs.tmp....2A} {p.~2}

\bibitem[\protect\citeauthoryear{{Blaes}, {Arras}  \& {Fragile}}{{Blaes}
  et~al.}{2006}]{2006BlaesArrasFragile}
{Blaes} O.~M.,  {Arras} P.,   {Fragile} P.~C.,  2006, \mn@doi [\mnras]
  {10.1111/j.1365-2966.2006.10370.x}, \href
  {https://ui.adsabs.harvard.edu/abs/2006MNRAS.369.1235B} {369, 1235}

\bibitem[\protect\citeauthoryear{{Breedt} et~al.,}{{Breedt}
  et~al.}{2010}]{2010BreedtNGC4051}
{Breedt} E.,  et~al., 2010, \mn@doi [\mnras]
  {10.1111/j.1365-2966.2009.16146.x}, \href
  {https://ui.adsabs.harvard.edu/abs/2010MNRAS.403..605B} {403, 605}

\bibitem[\protect\citeauthoryear{{Casebeer}, {Leighly}  \& {Baron}}{{Casebeer}
  et~al.}{2006}]{2006CasebeerREJ}
{Casebeer} D.~A.,  {Leighly} K.~M.,   {Baron} E.,  2006, \mn@doi [\apj]
  {10.1086/498125}, \href
  {https://ui.adsabs.harvard.edu/abs/2006ApJ...637..157C} {637, 157}

\bibitem[\protect\citeauthoryear{{Churazov}, {Gilfanov}  \&
  {Revnivtsev}}{{Churazov} et~al.}{2001}]{2001Churazov}
{Churazov} E.,  {Gilfanov} M.,   {Revnivtsev} M.,  2001, \mn@doi [\mnras]
  {10.1046/j.1365-8711.2001.04056.x}, \href
  {https://ui.adsabs.harvard.edu/abs/2001MNRAS.321..759C} {321, 759}

\bibitem[\protect\citeauthoryear{Coe}{Coe}{2009}]{coe2009fisher}
Coe D.,  2009, Fisher Matrices and Confidence Ellipses: A Quick-Start Guide and
  Software (\mn@eprint {arXiv} {0906.4123})

\bibitem[\protect\citeauthoryear{{Crummy}, {Fabian}, {Gallo}  \&
  {Ross}}{{Crummy} et~al.}{2006}]{2006CrummyCatalogue}
{Crummy} J.,  {Fabian} A.~C.,  {Gallo} L.,   {Ross} R.~R.,  2006, \mn@doi
  [\mnras] {10.1111/j.1365-2966.2005.09844.x}, \href
  {https://ui.adsabs.harvard.edu/abs/2006MNRAS.365.1067C} {365, 1067}

\bibitem[\protect\citeauthoryear{{Denney} et~al.,}{{Denney}
  et~al.}{2010}]{Denney2010}
{Denney} K.~D.,  et~al., 2010, \mn@doi [\apj] {10.1088/0004-637X/721/1/715},
  \href {https://ui.adsabs.harvard.edu/abs/2010ApJ...721..715D} {721, 715}

\bibitem[\protect\citeauthoryear{Done \& Jin}{Done \& Jin}{2016}]{2016DoneJin}
Done C.,  Jin C.,  2016, \mn@doi [Monthly Notices of the Royal Astronomical
  Society] {10.1093/mnras/stw1070}, 460, 1716

\bibitem[\protect\citeauthoryear{Done, Davis, Jin, Blaes  \& Ward}{Done
  et~al.}{2012a}]{Done2012}
Done C.,  Davis S.~W.,  Jin C.,  Blaes O.,   Ward M.,  2012a, \mn@doi [Monthly
  Notices of the Royal Astronomical Society]
  {10.1111/j.1365-2966.2011.19779.x}, 420, 1848

\bibitem[\protect\citeauthoryear{{Done}, {Davis}, {Jin}, {Blaes}  \&
  {Ward}}{{Done} et~al.}{2012b}]{2012DoneIntrinsicDiscEmission}
{Done} C.,  {Davis} S.~W.,  {Jin} C.,  {Blaes} O.,   {Ward} M.,  2012b, \mn@doi
  [\mnras] {10.1111/j.1365-2966.2011.19779.x}, \href
  {https://ui.adsabs.harvard.edu/abs/2012MNRAS.420.1848D} {420, 1848}

\bibitem[\protect\citeauthoryear{{Fabian} et~al.,}{{Fabian}
  et~al.}{2009}]{KLLinesin1H0707}
{Fabian} A.~C.,  et~al., 2009, \mn@doi [\nat] {10.1038/nature08007}, \href
  {https://ui.adsabs.harvard.edu/abs/2009Natur.459..540F} {459, 540}

\bibitem[\protect\citeauthoryear{Fragile, Blaes, Anninos  \& Salmonson}{Fragile
  et~al.}{2007}]{Fragile2007}
Fragile P.~C.,  Blaes O.~M.,  Anninos P.,   Salmonson J.~D.,  2007, \mn@doi
  [The Astrophysical Journal] {10.1086/521092}, 668, 417

\bibitem[\protect\citeauthoryear{Giacch\'e, Gilli  \& Titarchuk}{Giacch\'e
  et~al.}{2014}]{Giacch2014}
Giacch\'e S.,  Gilli R.,   Titarchuk L.,  2014, \mn@doi [Astronomy \&
  Astrophysics] {10.1051/0004-6361/201321904}, 562, A44

\bibitem[\protect\citeauthoryear{{Gierli{\'n}ski} \& {Done}}{{Gierli{\'n}ski}
  \& {Done}}{2004}]{2004GierlinskiDone}
{Gierli{\'n}ski} M.,  {Done} C.,  2004, \mn@doi [\mnras]
  {10.1111/j.1365-2966.2004.07266.x}, \href
  {https://ui.adsabs.harvard.edu/abs/2004MNRAS.347..885G} {347, 885}

\bibitem[\protect\citeauthoryear{{Gierlinski}, {Middleton}, {Ward}  \&
  {Done}}{{Gierlinski} et~al.}{2008}]{2008Nature}
{Gierlinski} M.,  {Middleton} M.,  {Ward} M.,   {Done} C.,  2008, \mn@doi
  [Nature] {10.1038/nature07277}, \href
  {http://ukads.nottingham.ac.uk/abs/2008Natur.455..369G} {455, 369}

\bibitem[\protect\citeauthoryear{{Gonz{\'a}lez-Mart{\'{\i}}n} \&
  {Vaughan}}{{Gonz{\'a}lez-Mart{\'{\i}}n} \&
  {Vaughan}}{2012}]{2012GonzMartinVaughan}
{Gonz{\'a}lez-Mart{\'{\i}}n} O.,  {Vaughan} S.,  2012, \mn@doi [\aap]
  {10.1051/0004-6361/201219008}, \href
  {http://ukads.nottingham.ac.uk/abs/2012A\%26A...544A..80G} {544, A80}

\bibitem[\protect\citeauthoryear{{Green}, {McHardy}  \& {Done}}{{Green}
  et~al.}{1999}]{1999GreenDoneNGC4051nonlinearvar}
{Green} A.~R.,  {McHardy} I.~M.,   {Done} C.,  1999, \mn@doi [\mnras]
  {10.1046/j.1365-8711.1999.02370.x}, \href
  {https://ui.adsabs.harvard.edu/abs/1999MNRAS.305..309G} {305, 309}

\bibitem[\protect\citeauthoryear{Hunter}{Hunter}{2007}]{Matplotlib}
Hunter J.~D.,  2007, \mn@doi [Computing in Science \& Engineering]
  {10.1109/MCSE.2007.55}, 9, 90

\bibitem[\protect\citeauthoryear{{Ingram} \& {Done}}{{Ingram} \&
  {Done}}{2011}]{2011IngramDone}
{Ingram} A.,  {Done} C.,  2011, \mn@doi [\mnras]
  {10.1111/j.1365-2966.2011.18860.x}, \href
  {https://ui.adsabs.harvard.edu/abs/2011MNRAS.415.2323I} {415, 2323}

\bibitem[\protect\citeauthoryear{{Ingram}, {van der Klis}, {Middleton}, {Done},
  {Altamirano}, {Heil}, {Uttley}  \& {Axelsson}}{{Ingram}
  et~al.}{2016}]{2016Ingram}
{Ingram} A.,  {van der Klis} M.,  {Middleton} M.,  {Done} C.,  {Altamirano} D.,
   {Heil} L.,  {Uttley} P.,   {Axelsson} M.,  2016, \mn@doi [\mnras]
  {10.1093/mnras/stw1245}, \href
  {https://ui.adsabs.harvard.edu/abs/2016MNRAS.461.1967I} {461, 1967}

\bibitem[\protect\citeauthoryear{{Jin}, {Done}  \& {Ward}}{{Jin}
  et~al.}{2020}]{2020JinREJ}
{Jin} C.,  {Done} C.,   {Ward} M.,  2020, arXiv e-prints, \href
  {https://ui.adsabs.harvard.edu/abs/2020arXiv200505857J} {p. arXiv:2005.05857}

\bibitem[\protect\citeauthoryear{{Kara}, {Fabian}, {Cackett}, {Steiner},
  {Uttley}, {Wilkins}  \& {Zoghbi}}{{Kara} et~al.}{2013}]{Kara20131H1Msdata}
{Kara} E.,  {Fabian} A.~C.,  {Cackett} E.~M.,  {Steiner} J.~F.,  {Uttley} P.,
  {Wilkins} D.~R.,   {Zoghbi} A.,  2013, \mn@doi [\mnras]
  {10.1093/mnras/sts155}, \href
  {https://ui.adsabs.harvard.edu/abs/2013MNRAS.428.2795K} {428, 2795}

\bibitem[\protect\citeauthoryear{Kara, Garc\'{i}a, Lohfink, Fabian, Reynolds,
  Tombesi  \& Wilkins}{Kara et~al.}{2017}]{Kara2017ARK564}
Kara E.,  Garc\'{i}a J.~A.,  Lohfink A.,  Fabian A.~C.,  Reynolds C.~S.,
  Tombesi F.,   Wilkins D.~R.,  2017, \mn@doi [Monthly Notices of the Royal
  Astronomical Society] {10.1093/mnras/stx792}, 468, 3489

\bibitem[\protect\citeauthoryear{Kara et~al.,}{Kara et~al.}{2019}]{Kara2019}
Kara E.,  et~al., 2019, \mn@doi [Nature] {10.1038/s41586-018-0803-x}, 565, 198

\bibitem[\protect\citeauthoryear{Liddle}{Liddle}{2007}]{Liddle2007}
Liddle A.~R.,  2007, \mn@doi [Monthly Notices of the Royal Astronomical
  Society: Letters] {10.1111/j.1745-3933.2007.00306.x}, 377, L74

\bibitem[\protect\citeauthoryear{{Markowitz}, {Papadakis}, {Ar{\'e}valo},
  {Turner}, {Miller}  \& {Reeves}}{{Markowitz}
  et~al.}{2007}]{2007MarkowitzMRK766}
{Markowitz} A.,  {Papadakis} I.,  {Ar{\'e}valo} P.,  {Turner} T.~J.,  {Miller}
  L.,   {Reeves} J.~N.,  2007, \mn@doi [\apj] {10.1086/510616}, \href
  {https://ui.adsabs.harvard.edu/abs/2007ApJ...656..116M} {656, 116}

\bibitem[\protect\citeauthoryear{{McHardy}, {Koerding}, {Knigge}, {Uttley}  \&
  {Fender}}{{McHardy} et~al.}{2006}]{2006McHardyAGNScaling}
{McHardy} I.~M.,  {Koerding} E.,  {Knigge} C.,  {Uttley} P.,   {Fender} R.~P.,
  2006, \mn@doi [\nat] {10.1038/nature05389}, \href
  {https://ui.adsabs.harvard.edu/abs/2006Natur.444..730M} {444, 730}

\bibitem[\protect\citeauthoryear{{McHardy}, {Ar{\'e}valo}, {Uttley},
  {Papadakis}, {Summons}, {Brinkmann}  \& {Page}}{{McHardy}
  et~al.}{2007}]{2007IMcHardyArk564Lor}
{McHardy} I.~M.,  {Ar{\'e}valo} P.,  {Uttley} P.,  {Papadakis} I.~E.,
  {Summons} D.~P.,  {Brinkmann} W.,   {Page} M.~J.,  2007, \mn@doi [\mnras]
  {10.1111/j.1365-2966.2007.12411.x}, \href
  {https://ui.adsabs.harvard.edu/abs/2007MNRAS.382..985M} {382, 985}

\bibitem[\protect\citeauthoryear{{Middleton} \& {Done}}{{Middleton} \&
  {Done}}{2010}]{2010MiddletonBHBAnalogy}
{Middleton} M.,  {Done} C.,  2010, \mn@doi [\mnras]
  {10.1111/j.1365-2966.2009.15969.x}, \href
  {http://ukads.nottingham.ac.uk/abs/2010MNRAS.403....9M} {403, 9}

\bibitem[\protect\citeauthoryear{{Middleton}, {Done}  \&
  {Gierli{\'n}ski}}{{Middleton} et~al.}{2007}]{2007MiddletonPGQSRs}
{Middleton} M.,  {Done} C.,   {Gierli{\'n}ski} M.,  2007, \mn@doi [\mnras]
  {10.1111/j.1365-2966.2007.12341.x}, \href
  {https://ui.adsabs.harvard.edu/abs/2007MNRAS.381.1426M} {381, 1426}

\bibitem[\protect\citeauthoryear{{Middleton}, {Done}, {Ward}, {Gierli{\'n}ski}
  \& {Schurch}}{{Middleton} et~al.}{2009}]{2009MiddletonSoftExcess}
{Middleton} M.,  {Done} C.,  {Ward} M.,  {Gierli{\'n}ski} M.,   {Schurch} N.,
  2009, \mn@doi [\mnras] {10.1111/j.1365-2966.2008.14255.x}, \href
  {http://ukads.nottingham.ac.uk/abs/2009MNRAS.394..250M} {394, 250}

\bibitem[\protect\citeauthoryear{{Middleton}, {Uttley}  \& {Done}}{{Middleton}
  et~al.}{2011}]{2011Middleton8Years}
{Middleton} M.,  {Uttley} P.,   {Done} C.,  2011, \mn@doi [\mnras]
  {10.1111/j.1365-2966.2011.19185.x}, \href
  {http://ukads.nottingham.ac.uk/abs/2011MNRAS.417..250M} {417, 250}

\bibitem[\protect\citeauthoryear{{Morgan}, {Remillard}  \& {Greiner}}{{Morgan}
  et~al.}{1997}]{1997MorganGRS1915}
{Morgan} E.~H.,  {Remillard} R.~A.,   {Greiner} J.,  1997, \mn@doi [\apj]
  {10.1086/304191}, \href
  {https://ui.adsabs.harvard.edu/abs/1997ApJ...482..993M} {482, 993}

\bibitem[\protect\citeauthoryear{{Motta}}{{Motta}}{2016}]{2016BHBQPOs}
{Motta} S.~E.,  2016, \mn@doi [Astronomische Nachrichten]
  {10.1002/asna.201612320}, \href
  {http://ukads.nottingham.ac.uk/abs/2016AN....337..398M} {337, 398}

\bibitem[\protect\citeauthoryear{{Nowak}}{{Nowak}}{1995}]{1995Nowak}
{Nowak} M.~A.,  1995, \mn@doi [\pasp] {10.1086/133679}, \href
  {https://ui.adsabs.harvard.edu/abs/1995PASP..107.1207N} {107, 1207}

\bibitem[\protect\citeauthoryear{{Pan}, {Yuan}, {Yao}, {Zhou}, {Liu}, {Zhou}
  \& {Zhang}}{{Pan} et~al.}{2016}]{2016Pan1HQPO}
{Pan} H.-W.,  {Yuan} W.,  {Yao} S.,  {Zhou} X.-L.,  {Liu} B.,  {Zhou} H.,
  {Zhang} S.-N.,  2016, \mn@doi [\apjl] {10.3847/2041-8205/819/2/L19}, \href
  {https://ui.adsabs.harvard.edu/abs/2016ApJ...819L..19P} {819, L19}

\bibitem[\protect\citeauthoryear{{Papadakis} \& {Lawrence}}{{Papadakis} \&
  {Lawrence}}{1995}]{1995PapadakisNGC4051qpo}
{Papadakis} I.~E.,  {Lawrence} A.,  1995, \mn@doi [\mnras]
  {10.1093/mnras/272.1.161}, \href
  {https://ui.adsabs.harvard.edu/abs/1995MNRAS.272..161P} {272, 161}

\bibitem[\protect\citeauthoryear{{Papadakis}, {Brinkmann}, {Negoro}  \&
  {Gliozzi}}{{Papadakis} et~al.}{2002}]{2002PapadakisArk564Break}
{Papadakis} I.~E.,  {Brinkmann} W.,  {Negoro} H.,   {Gliozzi} M.,  2002,
  \mn@doi [\aap] {10.1051/0004-6361:20011763}, \href
  {https://ui.adsabs.harvard.edu/abs/2002A&A...382L...1P} {382, L1}

\bibitem[\protect\citeauthoryear{{Pasham}, {Lin}, {Saxton}, {Jonker}, {Kara},
  {Stone}, {Maksym}  \& {Auchettl}}{{Pasham} et~al.}{2019a}]{2019PasahmTDEs}
{Pasham} D.,  {Lin} D.,  {Saxton} R.,  {Jonker} P.,  {Kara} E.,  {Stone} N.,
  {Maksym} P.,   {Auchettl} K.,  2019a, \baas, \href
  {https://ui.adsabs.harvard.edu/abs/2019BAAS...51c..27P} {51, 27}

\bibitem[\protect\citeauthoryear{{Pasham} et~al.,}{{Pasham}
  et~al.}{2019b}]{2019QPOsinTDEs}
{Pasham} D.~R.,  et~al., 2019b, \mn@doi [Science] {10.1126/science.aar7480},
  \href {https://ui.adsabs.harvard.edu/abs/2019Sci...363..531P} {363, 531}

\bibitem[\protect\citeauthoryear{Pedregosa et~al.,}{Pedregosa
  et~al.}{2011}]{scikit-learn}
Pedregosa F.,  et~al., 2011, Journal of Machine Learning Research, 12, 2825

\bibitem[\protect\citeauthoryear{{Peterson} et~al.,}{{Peterson}
  et~al.}{2000}]{2000PetersonNGC4051}
{Peterson} B.~M.,  et~al., 2000, \mn@doi [\apj] {10.1086/309518}, \href
  {https://ui.adsabs.harvard.edu/abs/2000ApJ...542..161P} {542, 161}

\bibitem[\protect\citeauthoryear{{Puchnarewicz} \& {Soria}}{{Puchnarewicz} \&
  {Soria}}{2002}]{2002PuchnarewiczREJ}
{Puchnarewicz} E.~M.,  {Soria} R.,  2002, arXiv e-prints, \href
  {https://ui.adsabs.harvard.edu/abs/2002astro.ph..2030P} {pp
  astro--ph/0202030}

\bibitem[\protect\citeauthoryear{Remillard \& McClintock}{Remillard \&
  McClintock}{2006}]{Remillard2006XRBproperties}
Remillard R.~A.,  McClintock J.~E.,  2006, \mn@doi [Annual Review of Astronomy
  and Astrophysics] {10.1146/annurev.astro.44.051905.092532}, 44, 49

\bibitem[\protect\citeauthoryear{Stella, Vietri  \& Morsink}{Stella
  et~al.}{1999}]{StellaVietri1999}
Stella L.,  Vietri M.,   Morsink S.~M.,  1999, \mn@doi [The Astrophysical
  Journal] {10.1086/312291}, 524, L63

\bibitem[\protect\citeauthoryear{{Timmer} \& {Koenig}}{{Timmer} \&
  {Koenig}}{1995}]{1995TK}
{Timmer} J.,  {Koenig} M.,  1995, \aap, \href
  {http://ukads.nottingham.ac.uk/abs/1995A\%26A...300..707T} {300, 707}

\bibitem[\protect\citeauthoryear{{Ueda}, {Yamaoka}  \& {Remillard}}{{Ueda}
  et~al.}{2009}]{2009UedaGRS1915}
{Ueda} Y.,  {Yamaoka} K.,   {Remillard} R.,  2009, \mn@doi [\apj]
  {10.1088/0004-637X/695/2/888}, \href
  {https://ui.adsabs.harvard.edu/abs/2009ApJ...695..888U} {695, 888}

\bibitem[\protect\citeauthoryear{{Uttley}, {McHardy}  \& {Papadakis}}{{Uttley}
  et~al.}{2002}]{2002UttleyBBN}
{Uttley} P.,  {McHardy} I.~M.,   {Papadakis} I.~E.,  2002, \mn@doi [\mnras]
  {10.1046/j.1365-8711.2002.05298.x}, \href
  {https://ui.adsabs.harvard.edu/abs/2002MNRAS.332..231U} {332, 231}

\bibitem[\protect\citeauthoryear{{Uttley}, {McHardy}  \& {Vaughan}}{{Uttley}
  et~al.}{2017}]{2017Uttleycoupling}
{Uttley} P.,  {McHardy} I.~M.,   {Vaughan} S.,  2017, \mn@doi [\aap]
  {10.1051/0004-6361/201630044}, \href
  {https://ui.adsabs.harvard.edu/abs/2017A&A...601L...1U} {601, L1}

\bibitem[\protect\citeauthoryear{{Vaughan}}{{Vaughan}}{2005}]{2005Vaughan}
{Vaughan} S.,  2005, \mn@doi [\aap] {10.1051/0004-6361:20041453}, \href
  {http://ukads.nottingham.ac.uk/abs/2005A\%26A...431..391V} {431, 391}

\bibitem[\protect\citeauthoryear{{Vaughan}}{{Vaughan}}{2010}]{2010BayesianVaughan}
{Vaughan} S.,  2010, \mn@doi [\mnras] {10.1111/j.1365-2966.2009.15868.x}, \href
  {http://ukads.nottingham.ac.uk/abs/2010MNRAS.402..307V} {402, 307}

\bibitem[\protect\citeauthoryear{Vaughan \& Uttley}{Vaughan \&
  Uttley}{2005}]{Vaughan05WhereAretheQPOs?}
Vaughan S.,  Uttley P.,  2005, \mn@doi [Monthly Notices of the Royal
  Astronomical Society] {10.1111/j.1365-2966.2005.09296.x}, 362, 235

\bibitem[\protect\citeauthoryear{{Vaughan}, {Edelson}, {Warwick}  \&
  {Uttley}}{{Vaughan} et~al.}{2003}]{2003UttleyVaughan}
{Vaughan} S.,  {Edelson} R.,  {Warwick} R.~S.,   {Uttley} P.,  2003, \mn@doi
  [\mnras] {10.1046/j.1365-2966.2003.07042.x}, \href
  {http://adsabs.harvard.edu/abs/2003MNRAS.345.1271V} {345, 1271}

\bibitem[\protect\citeauthoryear{{Vaughan}, {Uttley}, {Pounds}, {Nandra}  \&
  {Strohmayer}}{{Vaughan} et~al.}{2011}]{2011VaughanNGC4051}
{Vaughan} S.,  {Uttley} P.,  {Pounds} K.,  {Nandra} K.,   {Strohmayer} T.~E.,
  2011, \mn@doi [\mnras] {10.1111/j.1365-2966.2011.18319.x}, \href
  {https://ui.adsabs.harvard.edu/abs/2011MNRAS.413.2489V} {413, 2489}

\bibitem[\protect\citeauthoryear{{Vaughan}, {Uttley}, {Markowitz},
  {Huppenkothen}, {Middleton}, {Alston}, {Scargle}  \& {Farr}}{{Vaughan}
  et~al.}{2016}]{2016FalsePeriodicities}
{Vaughan} S.,  {Uttley} P.,  {Markowitz} A.~G.,  {Huppenkothen} D.,
  {Middleton} M.~J.,  {Alston} W.~N.,  {Scargle} J.~D.,   {Farr} W.~M.,  2016,
  \mn@doi [\mnras] {10.1093/mnras/stw1412}, \href
  {http://ukads.nottingham.ac.uk/abs/2016MNRAS.461.3145V} {461, 3145}

\bibitem[\protect\citeauthoryear{{Veron-Cetty} \& {Veron}}{{Veron-Cetty} \&
  {Veron}}{2010}]{2010VernonCetty}
{Veron-Cetty} M.~P.,  {Veron} P.,  2010, VizieR Online Data Catalog, \href
  {https://ui.adsabs.harvard.edu/abs/2010yCat.7258....0V} {p. VII/258}

\bibitem[\protect\citeauthoryear{{Virtanen} et~al.,}{{Virtanen}
  et~al.}{2020}]{Scipy}
{Virtanen} P.,  et~al., 2020, \mn@doi [Nature Methods]
  {https://doi.org/10.1038/s41592-019-0686-2}, \href {https://rdcu.be/b08Wh} {}

\bibitem[\protect\citeauthoryear{{Wang, T.} \& {Lu, Y.}}{{Wang, T.} \& {Lu,
  Y.}}{2001}]{Wang2001}
{Wang, T.} {Lu, Y.} 2001, \mn@doi [A\&A] {10.1051/0004-6361:20011071}, 377, 52

\bibitem[\protect\citeauthoryear{Wilkins \& Fabian}{Wilkins \&
  Fabian}{2013}]{Wilkins2013}
Wilkins D.~R.,  Fabian A.~C.,  2013, \mn@doi [Monthly Notices of the Royal
  Astronomical Society] {10.1093/mnras/sts591}, 430, 247

\bibitem[\protect\citeauthoryear{Zhang, Zhang, Yan, Fan  \& Liu}{Zhang
  et~al.}{2017}]{Zhang2017MRK766}
Zhang P.,  Zhang P.-f.,  Yan J.-z.,  Fan Y.-z.,   Liu Q.-z.,  2017, \mn@doi
  [The Astrophysical Journal] {10.3847/1538-4357/aa8d6e}, 849, 9

\bibitem[\protect\citeauthoryear{{Zhang}, {Zhang}, {Liao}, {Yan}, {Fan}  \&
  {Liu}}{{Zhang} et~al.}{2018}]{Zhang20181H}
{Zhang} P.-f.,  {Zhang} P.,  {Liao} N.-h.,  {Yan} J.-z.,  {Fan} Y.-z.,   {Liu}
  Q.-z.,  2018, \mn@doi [\apj] {10.3847/1538-4357/aaa29a}, \href
  {https://ui.adsabs.harvard.edu/abs/2018ApJ...853..193Z} {853, 193}

\bibitem[\protect\citeauthoryear{{van der Klis}}{{van der
  Klis}}{1988}]{1988VandK}
{van der Klis} M.,  1988, in Timing Neutron Stars, eds. H. Ogelman and E.P.J.
  van den Heuvel. NATO ASI Series C, Vol. 262, p. 27-70. Dordrecht: Kluwer,
  1988.. pp 27--70

\bibitem[\protect\citeauthoryear{{van der Walt}, {Colbert}  \&
  {Varoquaux}}{{van der Walt} et~al.}{2011}]{Numpy}
{van der Walt} S.,  {Colbert} S.~C.,   {Varoquaux} G.,  2011, \mn@doi
  [Computing in Science Engineering] {10.1109/MCSE.2011.37}, 13, 22

\makeatother
\end{thebibliography}




\begin{table*}

\caption{Details of the observations used in this study. The columns indicate: (1) Active Galactic Nuclei in the sample, (2): Observation ID, (3): The total EPIC-PN exposure time, (4): The length of the longest flare-less segment within the observation. The full table is available online.
}

\begin{tabular}{cccc}
  \hline
  AGN & Obs ID & Exposure Time (ks) & Maximum Flareless Exposure (ks) \\
  
 \hline
    
PG 0003+199 & 0101040101 & 31.60 & 23.44 \\
PG 0003+199 & 0306870101 & 132.70 & 132.70 \\
PG 0003+199 & 0510010701 & 16.70 & 16.70 \\
PG 0003+199 & 0600540501 & 80.70 & 34.96 \\
PG 0003+199 & 0600540601 & 130.40 & 130.40 \\
PG 0003+199 & 0741280201 & 137.40 & 137.40 \\
PG 0050+124 & 0110890301 & 20.00 & 20.00 \\
PG 0050+124 & 0300470101 & 82.90 & 24.00 \\
PG 0050+124 & 0743050301 & 20.40 & 2.37 \\
PG 0050+124 & 0743050801 & 20.40 & 20.40 \\
PG 0157+001 & 0101640201 & 11.20 & 10.19 \\

\end{tabular}

\label{tab:OBS}
\end{table*}




\bsp	
\label{lastpage}
\end{document}